\documentclass[sigplan,screen]{acmart}
\setcopyright{rightsretained}
\acmPrice{}
\acmDOI{10.1145/3453483.3454039}
\acmYear{2021}
\copyrightyear{2021}
\acmSubmissionID{pldi21main-p82-p}
\acmISBN{978-1-4503-8391-2/21/06}
\acmConference[PLDI '21]{Proceedings of the 42nd ACM SIGPLAN International Conference on Programming Language Design and Implementation}{June 20--25, 2021}{Virtual, Canada}
\acmBooktitle{Proceedings of the 42nd ACM SIGPLAN International Conference on Programming Language Design and Implementation (PLDI '21), June 20--25, 2021, Virtual, Canada}

\usepackage{booktabs}   
\usepackage{subcaption} 

\usepackage{balance}
\usepackage{enumitem}
\usepackage{hyperref}
\usepackage{listings}
\usepackage{xspace}
\usepackage{cleveref}
\usepackage{multirow}
\usepackage{microtype}
\usepackage{soul,xcolor}
\usepackage{mathpartir}
\usepackage{amsmath,amsthm}
\usepackage[utf8]{inputenc}
\DeclareUnicodeCharacter{2200}{\ensuremath{\forall}} 
\DeclareUnicodeCharacter{2083}{\ensuremath{_3}} 
\DeclareUnicodeCharacter{2102}{\ensuremath{\mathbb{C}}} 
\DeclareUnicodeCharacter{D7}{\ensuremath{\times}} 
\DeclareUnicodeCharacter{2287}{\ensuremath{\supseteq}} 
\DeclareUnicodeCharacter{2208}{\ensuremath{\in}} 
\DeclareUnicodeCharacter{230A}{\ensuremath{\lfloor{}}} 
\DeclareUnicodeCharacter{391}{\ensuremath{\Alpha}} 
\DeclareUnicodeCharacter{2293}{\ensuremath{\sqcap}} 
\DeclareUnicodeCharacter{1D502}{\ensuremath{\mathcal{y}}} 
\DeclareUnicodeCharacter{395}{\ensuremath{\Epsilon}} 
\DeclareUnicodeCharacter{2297}{\ensuremath{\otimes}} 
\DeclareUnicodeCharacter{399}{\ensuremath{\Iota}} 
\DeclareUnicodeCharacter{2218}{\ensuremath{\circ}} 
\DeclareUnicodeCharacter{229B}{\ensuremath{?}} 
\DeclareUnicodeCharacter{211A}{\ensuremath{\mathbb{Q}}} 
\DeclareUnicodeCharacter{39D}{\ensuremath{\Nu}} 
\DeclareUnicodeCharacter{3A1}{\ensuremath{\Rho}} 
\DeclareUnicodeCharacter{3A5}{\ensuremath{\Upsilon}} 
\DeclareUnicodeCharacter{22A7}{\ensuremath{\models}} 
\DeclareUnicodeCharacter{3A9}{\ensuremath{\Omega}} 
\DeclareUnicodeCharacter{2228}{\ensuremath{\lor}} 
\DeclareUnicodeCharacter{3B1}{\ensuremath{\alpha}} 
\DeclareUnicodeCharacter{B3}{\ensuremath{^3}} 
\DeclareUnicodeCharacter{3B5}{\ensuremath{\epsilon}} 
\DeclareUnicodeCharacter{3B9}{\ensuremath{\iota}} 
\DeclareUnicodeCharacter{3BD}{\ensuremath{\nu}} 
\DeclareUnicodeCharacter{1D53E}{\ensuremath{\mathbb{G}}} 
\DeclareUnicodeCharacter{3C1}{\ensuremath{\rho}} 
\DeclareUnicodeCharacter{22C3}{\ensuremath{\bigcup}} 
\DeclareUnicodeCharacter{1D542}{\ensuremath{\mathbb{K}}} 
\DeclareUnicodeCharacter{3C5}{\ensuremath{\upsilon}} 
\DeclareUnicodeCharacter{1D546}{\ensuremath{\mathbb{O}}} 
\DeclareUnicodeCharacter{3C9}{\ensuremath{\omega}} 
\DeclareUnicodeCharacter{1D54A}{\ensuremath{\mathbb{S}}} 
\DeclareUnicodeCharacter{1D54E}{\ensuremath{\mathbb{W}}} 
\DeclareUnicodeCharacter{1D4D3}{\ensuremath{\mathcal{D}}} 
\DeclareUnicodeCharacter{1D552}{\ensuremath{\mathbb{a}}} 
\DeclareUnicodeCharacter{1D4D7}{\ensuremath{\mathcal{H}}} 
\DeclareUnicodeCharacter{1D556}{\ensuremath{\mathbb{e}}} 
\DeclareUnicodeCharacter{1D4DB}{\ensuremath{\mathcal{L}}} 
\DeclareUnicodeCharacter{1D55A}{\ensuremath{\mathbb{i}}} 
\DeclareUnicodeCharacter{1D4DF}{\ensuremath{\mathcal{P}}} 
\DeclareUnicodeCharacter{1D55E}{\ensuremath{\mathbb{m}}} 
\DeclareUnicodeCharacter{1D4E3}{\ensuremath{\mathcal{T}}} 
\DeclareUnicodeCharacter{1D562}{\ensuremath{\mathbb{q}}} 
\DeclareUnicodeCharacter{2264}{\ensuremath{\leq}} 
\DeclareUnicodeCharacter{1D4E7}{\ensuremath{\mathcal{X}}} 
\DeclareUnicodeCharacter{1D566}{\ensuremath{\mathbb{u}}} 
\DeclareUnicodeCharacter{1D4EB}{\ensuremath{\mathcal{b}}} 
\DeclareUnicodeCharacter{1D56A}{\ensuremath{\mathbb{y}}} 
\DeclareUnicodeCharacter{1D4EF}{\ensuremath{\mathcal{f}}} 
\DeclareUnicodeCharacter{2070}{\ensuremath{^0}} 
\DeclareUnicodeCharacter{1D4F3}{\ensuremath{\mathcal{j}}} 
\DeclareUnicodeCharacter{2074}{\ensuremath{^4}} 
\DeclareUnicodeCharacter{1D4F7}{\ensuremath{\mathcal{n}}} 
\DeclareUnicodeCharacter{27F9}{\ensuremath{\Longrightarrow}} 
\DeclareUnicodeCharacter{1D4FB}{\ensuremath{\mathcal{r}}} 
\DeclareUnicodeCharacter{1D4FF}{\ensuremath{\mathcal{v}}} 
\DeclareUnicodeCharacter{1D501}{\ensuremath{\mathcal{x}}} 
\DeclareUnicodeCharacter{2080}{\ensuremath{_0}} 
\DeclareUnicodeCharacter{2203}{\ensuremath{\exists}} 
\DeclareUnicodeCharacter{2084}{\ensuremath{_4}} 
\DeclareUnicodeCharacter{2309}{\ensuremath{\rceil{}}} 
\DeclareUnicodeCharacter{210D}{\ensuremath{\mathbb{H}}} 
\DeclareUnicodeCharacter{392}{\ensuremath{\Beta}} 
\DeclareUnicodeCharacter{2115}{\ensuremath{\mathbb{N}}} 
\DeclareUnicodeCharacter{2294}{\ensuremath{\sqcup}} 
\DeclareUnicodeCharacter{396}{\ensuremath{\Zeta}} 
\DeclareUnicodeCharacter{2119}{\ensuremath{\mathbb{P}}} 
\DeclareUnicodeCharacter{39A}{\ensuremath{\Kappa}} 
\DeclareUnicodeCharacter{211D}{\ensuremath{\mathbb{R}}} 
\DeclareUnicodeCharacter{39E}{\ensuremath{\Xi}} 
\DeclareUnicodeCharacter{22A4}{\ensuremath{\top}} 
\DeclareUnicodeCharacter{2227}{\ensuremath{\land}} 
\DeclareUnicodeCharacter{3A6}{\ensuremath{\Phi}} 
\DeclareUnicodeCharacter{3B2}{\ensuremath{\beta}} 
\DeclareUnicodeCharacter{3B6}{\ensuremath{\zeta}} 
\DeclareUnicodeCharacter{1D539}{\ensuremath{\mathbb{B}}} 
\DeclareUnicodeCharacter{3BA}{\ensuremath{\kappa}} 
\DeclareUnicodeCharacter{1D53D}{\ensuremath{\mathbb{F}}} 
\DeclareUnicodeCharacter{3BE}{\ensuremath{\xi}} 
\DeclareUnicodeCharacter{1D541}{\ensuremath{\mathbb{J}}} 
\DeclareUnicodeCharacter{22C0}{\ensuremath{\bigland}} 
\DeclareUnicodeCharacter{2243}{\ensuremath{\simeq}} 
\DeclareUnicodeCharacter{3C6}{\ensuremath{\phi}} 
\DeclareUnicodeCharacter{1D54D}{\ensuremath{\mathbb{V}}} 
\DeclareUnicodeCharacter{1D4D0}{\ensuremath{\mathcal{A}}} 
\DeclareUnicodeCharacter{21D2}{\ensuremath{\Rightarrow}} 
\DeclareUnicodeCharacter{1D555}{\ensuremath{\mathbb{d}}} 
\DeclareUnicodeCharacter{1D4D4}{\ensuremath{\mathcal{E}}} 
\DeclareUnicodeCharacter{1D559}{\ensuremath{\mathbb{h}}} 
\DeclareUnicodeCharacter{1D4D8}{\ensuremath{\mathcal{I}}} 
\DeclareUnicodeCharacter{1D55D}{\ensuremath{\mathbb{l}}} 
\DeclareUnicodeCharacter{1D4DC}{\ensuremath{\mathcal{M}}} 
\DeclareUnicodeCharacter{1D561}{\ensuremath{\mathbb{p}}} 
\DeclareUnicodeCharacter{1D4E0}{\ensuremath{\mathcal{Q}}} 
\DeclareUnicodeCharacter{1D565}{\ensuremath{\mathbb{t}}} 
\DeclareUnicodeCharacter{1D4E4}{\ensuremath{\mathcal{U}}} 
\DeclareUnicodeCharacter{27E6}{\ensuremath{\llbracket}} 
\DeclareUnicodeCharacter{1D569}{\ensuremath{\mathbb{x}}} 
\DeclareUnicodeCharacter{1D4E8}{\ensuremath{\mathcal{Y}}} 
\DeclareUnicodeCharacter{1D4EC}{\ensuremath{\mathcal{c}}} 
\DeclareUnicodeCharacter{1D4F0}{\ensuremath{\mathcal{g}}} 
\DeclareUnicodeCharacter{1D4F4}{\ensuremath{\mathcal{k}}} 
\DeclareUnicodeCharacter{27F6}{\ensuremath{\longrightarrow}} 
\DeclareUnicodeCharacter{1D4F8}{\ensuremath{\mathcal{o}}} 
\DeclareUnicodeCharacter{207B}{\ensuremath{^{-}}} 
\DeclareUnicodeCharacter{1D4FC}{\ensuremath{\mathcal{s}}} 
\DeclareUnicodeCharacter{2081}{\ensuremath{_1}} 
\DeclareUnicodeCharacter{1D500}{\ensuremath{\mathcal{w}}} 
\DeclareUnicodeCharacter{2202}{\ensuremath{\partial}} 
\DeclareUnicodeCharacter{2A06}{\ensuremath{\bigsqcup}} 
\DeclareUnicodeCharacter{2308}{\ensuremath{\lceil{}}} 
\DeclareUnicodeCharacter{2291}{\ensuremath{\sqsubseteq}} 
\DeclareUnicodeCharacter{393}{\ensuremath{\Gamma}} 
\DeclareUnicodeCharacter{2295}{\ensuremath{\oplus}} 
\DeclareUnicodeCharacter{397}{\ensuremath{\Eta}} 
\DeclareUnicodeCharacter{39B}{\ensuremath{\Lambda}} 
\DeclareUnicodeCharacter{39F}{\ensuremath{\Omicron}} 
\DeclareUnicodeCharacter{3A3}{\ensuremath{\Sigma}} 
\DeclareUnicodeCharacter{22A5}{\ensuremath{\bot}} 
\DeclareUnicodeCharacter{2124}{\ensuremath{\mathbb{Z}}} 
\DeclareUnicodeCharacter{3A7}{\ensuremath{\Chi}} 
\DeclareUnicodeCharacter{2026}{\ensuremath{\dots}} 
\DeclareUnicodeCharacter{1F4A9}{\ensuremath{\LaTeX}} 
\DeclareUnicodeCharacter{222A}{\ensuremath{\cup}} 
\DeclareUnicodeCharacter{3B3}{\ensuremath{\gamma}} 
\DeclareUnicodeCharacter{3B7}{\ensuremath{\eta}} 
\DeclareUnicodeCharacter{B9}{\ensuremath{^1}} 
\DeclareUnicodeCharacter{1D538}{\ensuremath{\mathbb{A}}} 
\DeclareUnicodeCharacter{3BB}{\ensuremath{\lambda}} 
\DeclareUnicodeCharacter{1D53C}{\ensuremath{\mathbb{E}}} 
\DeclareUnicodeCharacter{3BF}{\ensuremath{\omicron}} 
\DeclareUnicodeCharacter{22C1}{\ensuremath{\biglor}} 
\DeclareUnicodeCharacter{1D540}{\ensuremath{\mathbb{I}}} 
\DeclareUnicodeCharacter{3C3}{\ensuremath{\sigma}} 
\DeclareUnicodeCharacter{22C5}{\ensuremath{\cdot}} 
\DeclareUnicodeCharacter{1D544}{\ensuremath{\mathbb{M}}} 
\DeclareUnicodeCharacter{3C7}{\ensuremath{\chi}} 
\DeclareUnicodeCharacter{1D54C}{\ensuremath{\mathbb{U}}} 
\DeclareUnicodeCharacter{1D4D1}{\ensuremath{\mathcal{B}}} 
\DeclareUnicodeCharacter{1D550}{\ensuremath{\mathbb{Y}}} 
\DeclareUnicodeCharacter{1D4D5}{\ensuremath{\mathcal{F}}} 
\DeclareUnicodeCharacter{1D554}{\ensuremath{\mathbb{c}}} 
\DeclareUnicodeCharacter{1D4D9}{\ensuremath{\mathcal{J}}} 
\DeclareUnicodeCharacter{1D558}{\ensuremath{\mathbb{g}}} 
\DeclareUnicodeCharacter{1D4DD}{\ensuremath{\mathcal{N}}} 
\DeclareUnicodeCharacter{1D55C}{\ensuremath{\mathbb{k}}} 
\DeclareUnicodeCharacter{1D4E1}{\ensuremath{\mathcal{R}}} 
\DeclareUnicodeCharacter{1D560}{\ensuremath{\mathbb{o}}} 
\DeclareUnicodeCharacter{1D4E5}{\ensuremath{\mathcal{V}}} 
\DeclareUnicodeCharacter{1D564}{\ensuremath{\mathbb{s}}} 
\DeclareUnicodeCharacter{27E7}{\ensuremath{\rrbracket}} 
\DeclareUnicodeCharacter{1D4E9}{\ensuremath{\mathcal{Z}}} 
\DeclareUnicodeCharacter{1D568}{\ensuremath{\mathbb{w}}} 
\DeclareUnicodeCharacter{1D4ED}{\ensuremath{\mathcal{d}}} 
\DeclareUnicodeCharacter{1D4F1}{\ensuremath{\mathcal{h}}} 
\DeclareUnicodeCharacter{2192}{\ensuremath{\rightarrow}} 
\DeclareUnicodeCharacter{1D4F5}{\ensuremath{\mathcal{l}}} 
\DeclareUnicodeCharacter{1D4F9}{\ensuremath{\mathcal{p}}} 
\DeclareUnicodeCharacter{207A}{\ensuremath{^{+}}} 
\DeclareUnicodeCharacter{1D4FD}{\ensuremath{\mathcal{t}}} 
\DeclareUnicodeCharacter{1D503}{\ensuremath{\mathcal{z}}} 
\DeclareUnicodeCharacter{2082}{\ensuremath{_2}} 
\DeclareUnicodeCharacter{2A05}{\ensuremath{\bigsqcap}} 
\DeclareUnicodeCharacter{2286}{\ensuremath{\subseteq}} 
\DeclareUnicodeCharacter{230B}{\ensuremath{\rfloor{}}} 
\DeclareUnicodeCharacter{2292}{\ensuremath{\sqsupseteq}} 
\DeclareUnicodeCharacter{394}{\ensuremath{\Delta}} 
\DeclareUnicodeCharacter{398}{\ensuremath{\Theta}} 
\DeclareUnicodeCharacter{39C}{\ensuremath{\Mu}} 
\DeclareUnicodeCharacter{3A0}{\ensuremath{\Pi}} 
\DeclareUnicodeCharacter{22A2}{\ensuremath{\vdash}} 
\DeclareUnicodeCharacter{3A4}{\ensuremath{\Tau}} 
\DeclareUnicodeCharacter{2229}{\ensuremath{\cap}} 
\DeclareUnicodeCharacter{3A8}{\ensuremath{\Psi}} 
\DeclareUnicodeCharacter{B2}{\ensuremath{^2}} 
\DeclareUnicodeCharacter{3B4}{\ensuremath{\delta}} 
\DeclareUnicodeCharacter{3B8}{\ensuremath{\theta}} 
\DeclareUnicodeCharacter{1D53B}{\ensuremath{\mathbb{D}}} 
\DeclareUnicodeCharacter{3BC}{\ensuremath{\mu}} 
\DeclareUnicodeCharacter{3C0}{\ensuremath{\pi}} 
\DeclareUnicodeCharacter{1D543}{\ensuremath{\mathbb{L}}} 
\DeclareUnicodeCharacter{22C2}{\ensuremath{\bigcap}} 
\DeclareUnicodeCharacter{2245}{\ensuremath{\cong}} 
\DeclareUnicodeCharacter{3C4}{\ensuremath{\tau}} 
\DeclareUnicodeCharacter{3C8}{\ensuremath{\psi}} 
\DeclareUnicodeCharacter{1D54B}{\ensuremath{\mathbb{T}}} 
\DeclareUnicodeCharacter{1D54F}{\ensuremath{\mathbb{X}}} 
\DeclareUnicodeCharacter{1D553}{\ensuremath{\mathbb{b}}} 
\DeclareUnicodeCharacter{1D4D2}{\ensuremath{\mathcal{C}}} 
\DeclareUnicodeCharacter{1D557}{\ensuremath{\mathbb{f}}} 
\DeclareUnicodeCharacter{1D4D6}{\ensuremath{\mathcal{G}}} 
\DeclareUnicodeCharacter{1D55B}{\ensuremath{\mathbb{j}}} 
\DeclareUnicodeCharacter{1D4DA}{\ensuremath{\mathcal{K}}} 
\DeclareUnicodeCharacter{1D55F}{\ensuremath{\mathbb{n}}} 
\DeclareUnicodeCharacter{1D4DE}{\ensuremath{\mathcal{O}}} 
\DeclareUnicodeCharacter{2261}{\ensuremath{\equiv}} 
\DeclareUnicodeCharacter{1D563}{\ensuremath{\mathbb{r}}} 
\DeclareUnicodeCharacter{1D4E2}{\ensuremath{\mathcal{S}}} 
\DeclareUnicodeCharacter{2265}{\ensuremath{\geq}} 
\DeclareUnicodeCharacter{1D567}{\ensuremath{\mathbb{v}}} 
\DeclareUnicodeCharacter{1D4E6}{\ensuremath{\mathcal{W}}} 
\DeclareUnicodeCharacter{1D56B}{\ensuremath{\mathbb{z}}} 
\DeclareUnicodeCharacter{1D4EA}{\ensuremath{\mathcal{a}}} 
\DeclareUnicodeCharacter{1D4EE}{\ensuremath{\mathcal{e}}} 
\DeclareUnicodeCharacter{1D4F2}{\ensuremath{\mathcal{i}}} 
\DeclareUnicodeCharacter{1D4F6}{\ensuremath{\mathcal{m}}} 
\DeclareUnicodeCharacter{1D4FA}{\ensuremath{\mathcal{q}}} 
\DeclareUnicodeCharacter{1D4FE}{\ensuremath{\mathcal{u}}} 

\usepackage{mathtools}
\usepackage{tikz}
\usetikzlibrary{calc,decorations.pathmorphing,shapes}
\usepackage{stmaryrd}
\usepackage{adjustbox}

\newcommand{\rot}[2][70]{\adjustbox{angle=#1}{\textbf{#2}}}

\crefformat{section}{§#2#1#3}

\newenvironment{nop}{}{}
\newenvironment{smathpar}{
\begin{nop}\small\begin{mathpar}}{
\end{mathpar}\end{nop}\ignorespacesafterend}

\definecolor{Bittersweet}{rgb}{1.0, 0.44, 0.37}
\definecolor{MidnightBlue}{rgb}{0.0, 0.2, 0.4}
\definecolor{BrightBlue}{rgb}{0.0, 0.2, 0.7}
\definecolor{byzantine}{rgb}{0.74, 0.2, 0.64}
\definecolor{caribbeangreen}{rgb}{0.0, 0.8, 0.6}

\lstnewenvironment{code}[1][]
    { 
      \lstset{
        basicstyle=\ttfamily\footnotesize,
        #1}%
      \csname lst@setfirstlabel\endcsname}
    { 
      \csname lst@savefirstlabel\endcsname}

\lstMakeShortInline[columns=fullflexible,keepspaces]|

\newcommand{\olam}[2]{\lambda^o #1. #2}

\newcommand{\clam}[2]{\lambda^c #1. #2}
\newcommand{\lam}[2]{\Lambda #1. #2}

\newcommand{\env}{\epsilon}
\newcommand{\envext}[3]{#1[#2 \mapsto #3]}

\newcommand{\oclos}[3]{\llparenthesis \olam{#1}{#2}, #3 \rrparenthesis}
\newcommand{\cclos}[3]{\llparenthesis \clam{#1}{#2}, #3 \rrparenthesis}
\newcommand{\clos}[3]{\llparenthesis \lam{#1}{#2}, #3 \rrparenthesis}
\newcommand{\kw}[1]{\text{\bf #1}}
\newcommand{\effval}[2]{\textsf{eff} ~#1 ~#2}
\newcommand{\exnval}[1]{\textsf{exn} ~#1}

\newcommand{\handle}[2]{\kw{match} ~#1 ~\kw{with} ~#2}
\newcommand{\throw}[2]{\kw{raise} ~#1 ~#2}
\newcommand{\perform}[2]{\kw{perform} ~#1 ~#2}

\newcommand{\caseval}[2]{\kw{return}~#1 \mapsto #2}
\newcommand{\caseexn}[3]{\kw{exception} ~#1 ~#2 \allowbreak \mapsto #3}
\newcommand{\caseeff}[4]{\kw{effect} ~#1 ~#2 ~#3 \mapsto #4}

\newcommand{\farg}[2]{\langle #1 ~#2 \rangle_{\text a}}
\newcommand{\ffun}[1]{\langle #1 \rangle_{\text f}}
\newcommand{\faritha}[3]{\langle #1 ~#2 ~#3 \rangle_{\text b1}}
\newcommand{\farithb}[2]{\langle #1 ~#2 \rangle_{\text b2}}

\newcommand{\kcons}{\lhd}
\newcommand{\fiber}{\varphi}
\newcommand{\fl}{\psi} 
\newcommand{\hc}{\eta} 

\newcommand{\cstack}{\gamma} 
\newcommand{\ostack}{\omega} 
\newcommand{\cstacka}[2]{\big \lceil #1, #2 \big \rceil_{\text c}} 
\newcommand{\ostacka}[2]{\big \lceil #1, #2 \big \rceil_{\text o}} 
\newcommand{\ostackemp}{\bullet}
\newcommand{\stack}{\sigma}

\newcommand{\term}{\tau}
\newcommand{\config}{\mathfrak{C}}
\newcommand{\configa}[3]{\|#1,#2,#3\|}

\newcommand{\ostep}{\xrightarrow{\text o}}
\newcommand{\cstep}{\xrightarrow{\text c}}
\newcommand{\step}{\rightarrow}

\begin{document}

\title{Retrofitting Effect Handlers onto OCaml}

\author{KC Sivaramakrishnan}
\affiliation{
  \institution{\small IIT Madras}            
	\city{\small Chennai}
	\country{\small India}
}
\email{kcsrk@cse.iitm.ac.in}          

\author{Stephen Dolan}
\affiliation{
  \institution{\small OCaml Labs}            
	\city{\small Cambridge}
	\country{\small UK}
}
\email{stephen.dolan@cl.cam.ac.uk}

\author{Leo White}
\affiliation{
  \institution{\small Jane Street}            
	\city{\small London}
	\country{\small UK}
}
\email{leo@lpw25.net}

\author{Tom Kelly}
\affiliation{
  \institution{\small OCaml Labs}            
	\city{\small Cambridge}
	\country{\small UK}
}
\email{tom.kelly@cantab.net}

\author{Sadiq Jaffer}
\affiliation{
  \institution{\small Opsian and OCaml Labs} 
	\city{\small Cambridge}
	\country{\small UK}
}
\email{sadiq@toao.com}

\author{Anil Madhavapeddy}
\affiliation{
  \institution{\small University of Cambridge and OCaml Labs}            
	\city{\small Cambridge}
	\country{\small UK}
}
\email{avsm2@cl.cam.ac.uk}          

\begin{abstract}
Effect handlers have been gathering momentum as a mechanism for modular
	programming with user-defined effects. Effect handlers allow for non-local
	control flow mechanisms such as generators, async/await, lightweight threads
	and coroutines to be composably expressed.  We present a design and evaluate
	a full-fledged efficient implementation of effect handlers for OCaml, an
	industrial-strength multi-paradigm programming language. Our implementation
	strives to maintain the backwards compatibility and performance profile of
	existing OCaml code. Retrofitting effect handlers onto OCaml is challenging
	since OCaml does not currently have any non-local control flow mechanisms
	other than exceptions. Our implementation of effect handlers for OCaml: {\em
	(i)}~imposes a mean 1\% overhead on a comprehensive macro benchmark suite
	that does not use effect handlers; {\em (ii)}~remains compatible with program
	analysis tools that inspect the stack; and {\em (iii)}~is efficient for new
	code that makes use of effect handlers.
\end{abstract}



\maketitle

\section{Introduction}

Effect handlers~\cite{Plotkin09} provide a modular foundation for user-defined
effects. The key idea is to separate the definition of the effectful operations
from their interpretations, which are given by \emph{handlers} of the effects.
For example,
\begin{lstlisting}
effect In_line : in_channel -> string
\end{lstlisting}
declares an \emph{effect} |In_line|, which is parameterised with an input
channel of type |in_channel|, which when \emph{performed} returns a |string|
value. A computation can perform the |In_line| effect without knowing how the
|In_line| effect is implemented. This computation may be enclosed by different
handlers that handle |In_line| differently. For example, |In_line| may be
implemented by performing a blocking read on the input channel or performing
the read asynchronously by offloading it to an event loop such as |libuv|,
without changing the computation. Thanks to the separation of effectful
operations from their implementation, effect handlers enable new approaches to
modular programming. Effect handlers are a generalisation of exception
handlers, where, in addition to the effect being handled, the handler is
provided with the delimited continuation~\cite{Danvy90} of the perform site.
This continuation may be used to resume the suspended computation later. This
enables non-local control-flow mechanisms such as resumable exceptions,
lightweight threads, coroutines, generators and asynchronous I/O to be
composably expressed.

One of the primary motivations to extend OCaml with effect handlers is to
natively support asynchronous I/O in order to express highly scalable
concurrent applications such as web servers in \emph{direct style} (as opposed
to using \emph{callbacks}). Many programming languages, including OCaml,
require non-local changes to source code in order to support asynchronous I/O,
often leading to a dichotomy between synchronous and asynchronous
code~\cite{colour}. For asynchronous I/O, OCaml developers typically use
libraries such as Lwt~\cite{lwt} and Async~\cite[\S18]{rwo}, where asynchronous
functions are represented as monadic computations. In these libraries, while
asynchronous functions can call synchronous functions directly, the converse is
not true. In particular, any function that calls an asynchronous function will
also have to be marked as asynchronous. As a result, large parts of the
applications using these libraries end up being in monadic form. Languages such
as GHC Haskell and Go provide lightweight threads, which avoids the dichotomy
between synchronous and asynchronous code. However, these languages bake-in the
lightweight thread implementation into the runtime system. With effect
handlers, asynchronous I/O can be implemented directly in OCaml as a library
without imposing a monadic form on the users.

There are many research languages and libraries built around effect
handlers~\cite{Leijen14,Biernacki20,Hillerstrom20,Pyro,Frank,Eff}. Unlike these
efforts, our goal is to retrofit effect handlers onto the OCaml programming
language, which has been in continuous use for the past 25 years in large
codebases including verification tools~\cite{everest,Coq}, mission critical
software systems~\cite{astree} and latency sensitive networked
applications~\cite{Madhavapeddy13}. OCaml is particularly favoured for its
competitive yet predictable performance, with a fast foreign-function interface
(FFI). It has excellent compatibility with program analysis tools such as
debuggers and profilers that utilise DWARF stack unwind tables~\cite{DWARF} to
obtain a backtrace.

OCaml currently does not support any non-local control flow mechanisms other
than exceptions. This makes it particularly challenging to implement the
delimited continuations necessary for effect handlers without sacrificing the
desirable properties of OCaml. A standard way of implementing continuations is
to use continuation-passing style (CPS) in the compiler's intermediate
representation (IR)~\cite{Leijen14}. OCaml does not use a CPS IR, and changing
the compiler to utilise a CPS IR would be an enormous undertaking that would
affect the performance profile of existing OCaml applications due to the
increased memory allocations as the continuation closures get allocated on the
heap~\cite{Farvardin20}. Moreover, with CPS, an explicit stack is absent, and
hence, we would lose compatibility with tools that inspect the program stack.
Hence, we choose not to use CPS translation and represent the continuations as
call stacks.

The search for an expressive effect system that guarantees that all the effects
performed in the program are handled (\emph{effect safety}) in the presence of
advanced features such as polymorphism, modularity and generativity is an
active area of research~\cite{Leijen14, Biernacki19, Biernacki20,
Hillerstrom20}. We do not focus on this question in this paper, and our
implementation of effect handlers in OCaml does not guarantee effect safety. We
leave the question of effect safety for future work.

\vspace{-1mm}
\subsection{Requirements}
\label{sec:req}

We motivate our effect handler design based on the following ideal
requirements:

\begin{enumerate}[label=R\arabic*]
	\item \textbf{Backwards compatibility.} Existing OCaml programs do not break
		under OCaml extended with effect handlers. OCaml code that does not use
		effect handlers will pay minimal performance and memory cost.

	\item \textbf{Tool compatibility.} OCaml programs with effect handlers
		produce well-formed backtraces and remain compatible with program analysis
		tools such as debuggers and profilers that inspect the stack using DWARF
		unwind tables.

	\item \textbf{Effect handler efficiency.} The program must accommodate
		millions of continuations at the same time to support highly-concurrent
		applications. Installing effect handlers, capturing and resuming
		continuations must be fast.

	\item \textbf{Forwards compatibility.} As a cornerstone of modularity, we
		also want blocking I/O code to \emph{transparently} be made asynchronous
		with the help of effect handlers.
\end{enumerate}

The need to host millions of continuations at the same time rules out the use
of a large contiguous stack space as in C for continuations. Instead, we resort
to using small initial stacks and growing the stacks on demand. As a result,
OCaml functions, irrespective of whether they use effect handlers, need to
perform stack overflow checks, and external C functions (which do not have
stack overflow checks) must be performed on a separate system stack.
Additionally, we must generate DWARF stack unwind tables for stacks that may be
non-contiguous. In this work, we develop the compiler and runtime support
required for implementing efficient effect handlers for OCaml that satisfy
these requirements.

Our work is also timely. The WebAssembly~\cite{Hass17} community group is
considering effect handlers as one of the mechanisms for supporting
concurrency, asynchronous I/O and generators~\cite{WasmProposal}. Project
Loom~\cite{loom} is an OpenJDK project that adds virtual threads and delimited
continuations to Java. The Swift roadmap~\cite{swift} includes direct style
asynchronous programming and structured concurrency as milestones. We believe
that our design choices will inform similar choices to be made in other
industrial-strength languages.

\subsection{Contributions}

Our contributions are to present:

\begin{itemize}
	\item the design and implementation of effect handlers for OCaml. Our design
		retains OCaml's compatibility with program analysis tools that inspect the
		stack using DWARF unwind tables. We have validated our DWARF unwind tables
		with the assistance of an automated validator tool~\cite{Bastian19}.
	\item a formal operational semantics for the effect handler implementation in
		OCaml. Our formalism explicitly models the interactions with the C stack,
		which is generally overlooked by other formal models, but which the
		implementations must handle.
	\item extensive evaluation which shows that our implementation has minimal
		impact on code that does not use effect handlers, and serves as an
		efficient foundation for scalable concurrent programming.
\end{itemize}

We have implemented effect handlers in a multicore extension of the OCaml
programming language which we call \textbf{Multicore OCaml} to distinguish it
from \emph{stock OCaml}. Multicore OCaml delineates concurrency (overlapped
execution of tasks) from parallelism (simultaneous execution of tasks) with
distinct mechanisms for expressing them. Sivaramakrishnan et
al.~\cite{Sivaramakrishnan20} describe the parallelism support in Multicore
OCaml enabled by \emph{domains}. The focus of this paper is the concurrency
support enabled by effect handlers.

The remainder of the paper continues with a description of the stock OCaml
program stack (\S\ref{sec:stack}). We then describe effect handlers in
Multicore OCaml focussing on the challenges in retrofitting them into a
mainstream \emph{systems} language (\S\ref{sec:refine}), followed by the static
and dynamic semantics for Multicore OCaml effect handlers
(\S\ref{sec:semantics}). We then discuss the compiler and the runtime system
support for implementing effect handlers (\S\ref{sec:impl}), and present an
extensive performance evaluation of effect handlers (\S\ref{sec:eval}) against
our design goals (\S\ref{sec:req}). Finally, we discuss the related work
(\S\ref{sec:related}) and conclude (\S\ref{sec:conc}).

\section{Background: OCaml Stacks}
\label{sec:stack}

\begin{figure*}
\begin{minipage}{0.30\linewidth}
  \begin{minipage}{\linewidth}
		\begin{lstlisting}[xleftmargin=0.05\linewidth]
external ocaml_to_c
  : unit -> int = "ocaml_to_c"
exception E1
exception E2
let c_to_ocaml () = raise E1
let _ = Callback.register
    "c_to_ocaml" c_to_ocaml
let omain () =
  try (* h1 *)
    try (* h2 *) ocaml_to_c ()
    | with E2 -> 0
  | with E1 -> 42;;
let _ = assert (omain () = 42)
    \end{lstlisting}
    \vspace{-2mm}
    \subcaption{\texttt{meander.ml}}
    \label{code:meander_ml}
  \end{minipage}
  \begin{minipage}{\linewidth}
		\begin{lstlisting}[language=c, xleftmargin=0.05\linewidth]
#include <caml/mlvalues.h>
#include <caml/callback.h>

value ocaml_to_c (value unit) {
  caml_callback(*caml_named_value
    ("c_to_ocaml"), Val_unit);
  return Val_int(0);
}
    \end{lstlisting}
    \vspace{-4mm}
    \subcaption{\texttt{meander.c}}
    \label{code:meander_c}
  \end{minipage}
\end{minipage}
\begin{minipage}{0.30\linewidth}
  \centering
  \includegraphics[scale=0.44]{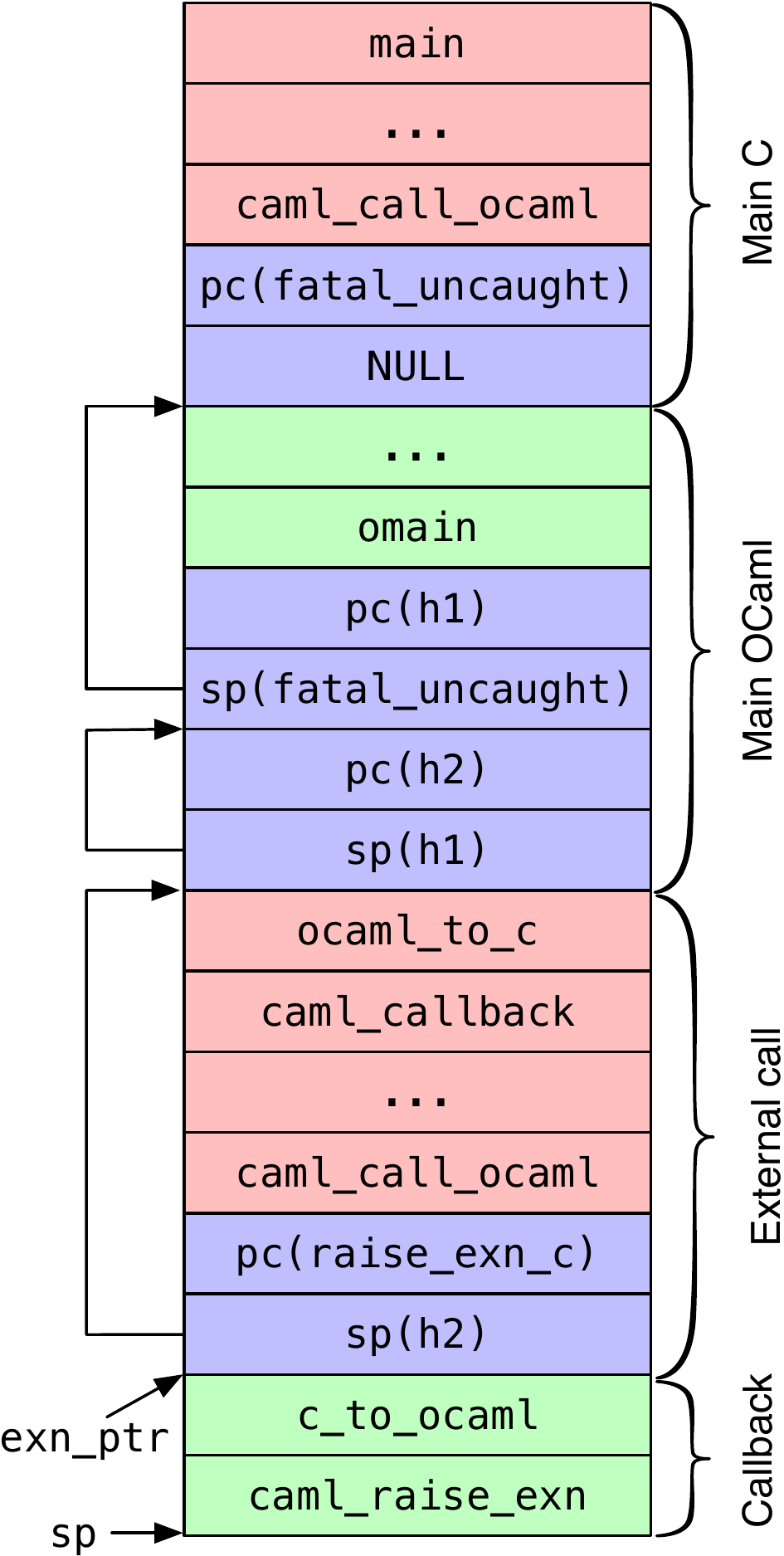}
  \subcaption{Stack layout before raise \texttt{E1}.}
  \label{fig:stock_stack}
\end{minipage}
\begin{minipage}{0.39\linewidth}
\begin{lstlisting}[language=c,basicstyle=\ttfamily\footnotesize]
#0  0x925dc in caml_raise_exn ()
#1  0x6fd3e in camlMeander__c_to_ocaml_83 () at meander.ml:5
#2  0x925a4 in caml_call_ocaml ()
#3  0x8a84a in caml_callback_exn (...) at callback.c:145
#4  caml_callback (...) at callback.c:199
#5  0x76e0a in ocaml_to_c (unit=1) at meander.c:5
#6  0x6fd77 in camlMeander__omain_88 () at meander.ml:10
#7  0x6fe92 in camlMeander__entry () at meander.ml:13
#8  0x6f719 in caml_program ()
#9  0x925a4 in caml_call_ocaml ()
#10 0x92e4c in caml_startup_common (...) at startup_nat.c:162
#11 0x92eab in caml_startup_exn (...) at startup_nat.c:167
#12 caml_startup (...) at startup_nat.c:172
#13 0x6f55c in main (...) at main.c:44
\end{lstlisting}
\subcaption{\texttt{gdb} backtrace before raise \texttt{E1}.}
\label{code:gdb_backtrace}
\end{minipage}
\vspace{-2mm}
\caption{Program stack on stock OCaml.}
\vspace{-2mm}
\end{figure*}

The main challenge in implementing effect handlers in Multicore OCaml is
managing the program stack and preserving its desirable properties. In this
section, we provide an overview of the program stack and related mechanisms in
stock OCaml.

Consider the layout of the stock OCaml stack for the program shown in
Figures~\ref{code:meander_ml} and~\ref{code:meander_c}. The OCaml main function
|omain| installs two exception handlers |h1| and |h2| to handle the exceptions
|E1| and |E2|. |omain| calls the external C function |ocaml_to_c|, which in
turn calls back into the OCaml function |c_to_ocaml|, which raises the
exception |E1|. OCaml supports raising exceptions in C as well as throwing
exceptions across external calls. Hence, the exception |E1| gets caught in the
handler |h1|, and |omain| returns |42|. The layout of the stack in the native
code backend just before raising the exception in |c_to_ocaml| is illustrated
in Figure~\ref{fig:stock_stack}. Note that the stack grows downwards.

OCaml uses the same program stack as C, and hence the stack has
alternating sequences of C and OCaml frames. However, unlike C, OCaml does not
create pointers into OCaml frames. OCaml uses the hardware support for |call|
and |return| instructions for function calls and returns. OCaml does not
perform explicit stack overflow checks in code, and, just like C, relies on the
guard page at the end of the stack region to detect stack overflow. Stack
overflow is detected by a memory fault and a |Stack_overflow| exception is
raised to unwind the stack.

\vspace{-1mm}
\subsection{External calls and callbacks}
\label{sec:external}

OCaml does not use the C calling convention. In particular, there are no
callee-saved registers in OCaml. In the |x86-64| backend, the OCaml runtime
makes use of two C callee-saved registers for supporting OCaml execution. The
register |r15| holds the \emph{allocation pointer} into the minor heap used for
bump pointer allocation, and |r14| holds a reference to the |Caml_state|, a
table of global variables used by the runtime. This makes external calls
extremely fast in OCaml. If the external function does not allocate in the
OCaml heap, then it can be called directly and no bookkeeping is necessary. For
external functions which allocate in the OCaml heap, the cached allocation
pointer is saved to |Caml_state| before the external call and it is restored on
return. Similarly, callbacks into OCaml from C are also cheap: these involve
loading the arguments in the right registers and calling the OCaml function.
OCaml callbacks are relatively common as the garbage collector (GC), which is
implemented in C, executes OCaml finalisation functions as callbacks.

\vspace{-1mm}
\subsection{Exception handlers}
\label{sec:exn_handlers}

The lack of callee-saved registers also makes exception handling fast. In the
absence of callee-saved registers, no registers need to be saved when entering
a |try| block. Similarly, no registers need to be restored when handling an
exception. Installing an exception handler simply pushes the program counter
(|pc|) of the handler and the current exception pointer (|exn_ptr| -- a field
in |Caml_state|) onto the stack. After this, the current exception pointer is
updated to be the current stack pointer (|rsp|). This creates a linked-list of
exception handler frames on the stack as shown in Figure~\ref{fig:stock_stack}.
Raising an exception simply sets |rsp| to |exn_ptr|, loads the saved |exn_ptr|,
and jumps to the |pc| of the handler.

In order to forward exceptions across C frames, the C stub function
|caml_call_ocaml|, pushes an exception handler frame that either forwards the
exception to the innermost OCaml exception handler (|raise_exn_c| in
Figure~\ref{fig:stock_stack}) or prints a fatal error (|fatal_uncaught|) if
there are no enclosing handlers. Exceptions are so cheap in OCaml that it is
common to use them for \emph{local} control flow.

\vspace{-1mm}
\subsection{Stack unwinding}
\label{sec:unwind}

OCaml generates \emph{stack maps} in order to accurately identify roots on the
stack for assisting the GC. For every call point in the program, the OCaml
compiler emits the size of the frame and the set of all live registers in the
frame that point to the heap. During a GC, the OCaml stack is walked and the
roots are marked, skipping over the C frames.

OCaml also generates precise DWARF unwind information for OCaml, thanks to
which debuggers such as |gdb| and |lldb|, and profilers such as |perf| work
out-of-the-box. For example, for the program in Figures~\ref{code:meander_ml}
and~\ref{code:meander_c}, one could set a break point in |gdb| at
|caml_raise_exn| to get the backtrace in Figure~\ref{code:gdb_backtrace} which
corresponds to the stack in Figure~\ref{fig:stock_stack}.

The same backtrace can also be obtained by using \emph{frame pointers} instead
of DWARF unwind tables. OCaml allows compiling code with frame pointers, but
they are not enabled by default. The OCaml stack tends to be deep with small
frames due to the pervasive use of recursive functions, not all of which are
tail-recursive. Hence, the addition of frame pointers can significantly
increase the size of the
stack\footnote{https://github.com/ocaml/ocaml/issues/5721\#issuecomment-472965549}.
Moreover, not using frame pointers saves two instructions in the function
prologue and epilogue, and makes an extra register (|rbp| on x86\_64)
available. Note that the DWARF unwind information is complementary to the
information used by OCaml to walk the stack for GC.

\vspace{-1mm}
\section{Effect Handlers}
\label{sec:refine}

In this section, we describe the effect handlers in Multicore OCaml, and refine
the design to retrofit them onto OCaml.

\vspace{-1mm}
\subsection{Asynchronous I/O}
\label{sec:async_io}

Since our primary motivation is to enable composable asynchronous I/O, let us
implement a cooperative lightweight thread library with support for forking new
threads and yielding control to other threads. We will then extend this library
with support for synchronously reading from channels and subsequently make it
asynchronous \textit{without changing the client code for asynchrony}. In order
to support forking and yielding threads, we declare the following effects:

\begin{lstlisting}
effect Fork : (unit -> unit) -> unit
effect Yield : unit
\end{lstlisting}

The |Fork| effect takes a thunk which is spawned as a concurrent thread, and
the |Yield| effect yields control to another thread in the scheduler queue. We
can define helper functions to perform these effects:

\begin{lstlisting}
let fork f = perform (Fork f)
let yield () = perform Yield
\end{lstlisting}

The implementation of the scheduler queue is defined in the |run| function,
which handles the effects appropriately:

\begin{lstlisting}[numbers=left, xleftmargin=0.025\textwidth]
let run main =
  let runq = Queue.create () in
  let suspend k = Queue.push k runq in
  let rec run_next () =
    match Queue.pop runq with
    | k -> continue k ()
    | exception Queue.Empty -> ()
  in
  let rec spawn f =
    match f () with
    | () -> run_next () (* value case *)
    | effect Yield k -> suspend k; run_next ()
    | effect (Fork f') k -> suspend k; spawn f'
  in
  spawn main
\end{lstlisting}

The function |spawn| (line 9) evaluates the computation |f| in an effect
handler. The computation |f| may return normally with a value, or perform
effects |Fork f'| and |Yield|. The pattern |effect Yield k| handles the effect
|Yield| and binds |k| to the continuation of the corresponding |perform|
delimited by this handler. The scheduler queue |runq| maintains a queue of
these continuations. |suspend| pushes continuations into the queue, |run_next|
pops continuations from the queue and resumes them with |()| value using the
|continue| primitive. In the case of the |Yield| effect, we suspend the current
continuation |k| and resume the next available continuation. In the case of the
|Fork f'| effect, we suspend the current continuation and recursively call
|spawn| on |f'| in order to run |f'| concurrently. Observe that we can change
the scheduling algorithm from FIFO to LIFO by changing the scheduler queue to a
stack.

We can implement support for synchronous read from channels by adding the
following case to the effect handler in |spawn|:

\begin{lstlisting}
let rec spawn f =
  match f () with
	...
  | effect (In_line ic) k -> continue k (input_line ic)
\end{lstlisting}
This uses OCaml's standard |input_line| function to read a line
synchronously from the channel |ic| and resume the continuation |k| with the
resultant string. However, performing reads synchronously blocks the entire
scheduler, preventing other threads from running until the I/O is completed.

We can make the I/O asynchronous by modifying the |run| function as follows:

\begin{lstlisting}[numbers=left, xleftmargin=.025\textwidth]
let run main =
  let runq = Queue.create () in
  let suspend k = Queue.push (continue k) runq in
  let pending_reads = ref [] in
  let rec run_next () =
    match Queue.pop runq with
    | f -> f ()
    | exception Queue.Empty ->
      match !pending_reads with
      | [] -> () (* no pending reads *)
      | todo ->
        let compl,todo = do_reads todo in
        List.iter (fun (s,k) ->
         Queue.push (fun () -> continue k s) runq) compl;
        pending_reads := todo;
        run_next ()
  in
  let rec spawn f =
    match f () with
    | () -> run_next () (* value case *)
    | effect Yield k -> suspend k; run_next ()
    | effect (Fork f') k -> suspend k; spawn f'
    | effect (In_line ic) k ->
      pending_reads := (ic,k)::!pending_reads;
      run_next ()
  in
  spawn main
\end{lstlisting}

The scheduler queue |runq| now holds thunks instead of continuations. The value
|pending_reads| maintains a list of pending reads and the associated
continuations (line 4). At line 24, we handle the |In_line| effect by pushing
the pair of input channel |ic| and continuation |k| to |pending_reads|,
allowing other threads in the scheduler to run.

When the scheduler queue is empty, the |run_next| function performs the pending
reads. We abstract away the details of the event-based I/O using the |do_reads|
function (line 12). |do_reads| takes a list of pending reads and blocks until
at least one of the reads succeeds. It returns a pair of lists |compl| and
|todo|. |compl| contains the result strings from successful reads and
corresponding continuations. These continuations are arranged to be resumed
with the read result and pushed into the scheduler queue. |todo| contains the
channels on which input is still pending and their corresponding continuations.
|pending_reads| is updated to point to the |todo| list so that they may be
attempted later. Observe that all of the changes to add asynchrony are
localised to the |run| function, and the computation that performs these
effects can remain in direct style (as opposed to the monadic-style in
Lwt and Async).

This example does not resume a continuation more than once. This also holds
true for other use cases such as generators and coroutines. Hence, our
continuations are one-shot, and resuming the continuation more than once raises
an |Invalid_argument| exception. It is well-known that one-shot continuations
can be implemented efficiently~\cite{Bruggeman96}.

While OCaml permits throwing exceptions across C frames, we do not allow
effects to propagate across C frames as the C frames would become part of the
captured continuation. Managing C frames as part of the continuation is a
complex endeavour~\cite{Leijen17}, and we find that the complexity budget
outweighs the relatively fewer mechanisms enabled by this addition in our
setting.

\subsection{Resource cleanup}

The interaction of non-local control flow with systems programming is quite
subtle~\cite{TFP17, Leijen18}. Consider the following function that uses
blocking I/O functions from the OCaml standard library to copy data from the
input channel |ic| to the output channel |oc|:
\begin{lstlisting}
let copy ic oc =
  let rec loop () =
	  output_string oc ((input_line ic) ^ "\n"); loop () in
	try loop () with
  | End_of_file -> close_in ic; close_out oc
	| e -> close_in ic; close_out oc; raise e
\end{lstlisting}
The function |input_line| raises an |End_of_file| exception on reaching the end
of input, which is handled by the exception handler which closes the channels.
The |close_*| functions do nothing if the channel is already closed. The code
is written in a defensive style to handle other exceptional cases such as the
channels being closed externally. Both |input_line| and |output_string| raise
a |Sys_error| exception if the channel is closed. In this case, the catch-all
exception handler closes the channels and reraises the exception to communicate
the exceptional behaviour to the caller.

One of our goals (\S\ref{sec:req}) is to make this code transparently
asynchronous. We can define effects for performing the I/O operations and wrap
them up in functions with the same signature as the one from the standard
library:
\begin{lstlisting}
effect In_line : in_channel -> string
effect Out_str : out_channel * string -> unit
let input_line ic = perform (In_line ic)
let output_string oc s = perform (Out_str (oc, s))
\end{lstlisting}

We can then use the |run| function that we defined earlier, to discharge the
I/O operations asynchronously and resume with the result. While this handles
value return cases, what about the exceptional cases |End_of_file| and
|Sys_error|? To this end, we introduce a |discontinue| primitive to resume a
continuation by raising an exception. In this example, on reaching the end of
file, we would discontinue the captured continuation of the |input_line|
function with |discontinue k End_of_file|, which raises the exception at
|input_line| call site, and the open channels will be closed.

OCaml programs that use resources such as channels are usually written
defensively with the assumption that calling a function will return
\emph{exactly once}, either normally or exceptionally. Since effect handlers in
Multicore OCaml do not ensure that all the effects are handled, if the function
performs an effect with no matching handler, then the function \emph{will not
return at all}. To remedy this, when such an effect bubbles up to the
top-level, we discontinue the continuation with an |Unhandled| exception so
that the exception handlers may run and clean up the resources.

\vspace{-1mm}
\section{Semantics}
\label{sec:semantics}

In this section, we formalise the effect handler design for Multicore OCaml.

\vspace{-1mm}
\subsection{Static semantics}
\label{sec:static_semantics}

As mentioned earlier, effect handlers in Multicore OCaml do not guarantee
effect safety. Programs without matching effect handlers are well-typed
Multicore OCaml programs. As a result, our static semantics is simpler than
languages that ensure effect safety~\cite{Eff, Hillerstrom20, Leijen14, Effekt,
Frank, Biernacki19}. This is important for backwards compatibility as our goal
is to retrofit effect handlers to a language with large legacy codebases;
programs that do not use effects remain well-typed, and those that do compose
well with those that don't.

The static semantics of effect handlers in OCaml is captured succinctly by its
API:
\begin{lstlisting}
type 'a eff = ..
type ('a,'b) continuation
val perform: 'a eff -> 'a
val continue: ('a,'b) continuation -> 'a -> 'b
val discontinue: ('a,'b) continuation -> exn -> 'b
(* Internal API *)
type 'a comp = unit -> 'a
type ('a,'b) handler =
{retc: 'a -> 'b;
 effc: 'c.'c eff -> ('c,'b) continuation -> 'b; }
val match_with: 'a comp -> ('a,'b) handler -> 'b
\end{lstlisting}
We introduce an extensible variant type~\cite{ExtVariants} |'a eff| of effect
values, which when performed using the |perform| primitive returns an |'a|
value. Constructors for the value of type |'a eff| are declared using the
effect declarations. For example, the declaration |effect E : string -> int|
is syntactic sugar for adding a new constructor to the variant type
|type _ eff += E : string -> int eff|. We introduce the type
|('a,'b) continuation| of delimited continuations which expects a |'a| value for
resumption and returns a |'b| value. The continuations may be continued with a
suitably typed value or discontinued with an exception.

For effect handling, we extend OCaml's |match ... with| syntax with effect
patterns. The expression
\begin{lstlisting}
match e with
| None -> false | Some b -> b
| effect (E s) k1 -> e1 | effect (F f) k2 -> e2
\end{lstlisting}
\noindent is translated to the equivalent of
\begin{lstlisting}
match_with (fun () -> e)
{ retc = (function None -> false | Some b -> b);
  effc = (function
  | (E s) -> (fun k1 -> e1)
  | (F f) -> (fun k2 -> e2)
  | e -> (fun k -> match perform e with
          | v -> continue k v
          | exception e -> discontinue k e)); }
\end{lstlisting}
For the sake of exposition, we introduce a |('a,'b) handler| type. This handler
handles a |'a comp| that returns a |'a| value, and itself returns a |'b| value.
The handler has a return field |retc| of type |'a -> 'b|. The effect field
|effc| handles effects of type |'c eff| with |('c,'b) continuation| and returns
a value of type |'b|. The last case in |effc| \emph{reperforms} any unmatched
effect to the outer handler and returns the value and exceptions back to the
original performer. In the implementation, reperform is implemented as a
primitive to avoid executing code on the resumption path.

\begin{figure*}
\begin{minipage}{\linewidth}
  \begin{minipage}[t]{0.49\linewidth}
  \begin{smathpar}
  \begin{array}{rrcl}
    \text{Constants} & n & \coloneqq & ℤ \\
    \text{Abstractions} & \Lambda & \coloneqq & \lambda^o \mid \lambda^c \\
    \text{Expressions} & e & \coloneqq & n \mid x \mid e ~e \mid  \lam{x}{e} \mid e \odot e \mid \throw{l}{e} \\
                       &   & \mid      & \handle{e}{h} \mid \perform{l}{e} \\
    \text{Handlers} & h & \coloneqq & \{\caseval{x}{e}\} \mid \{\caseexn{l}{x}{e}\} \uplus h \\
                   &   & \mid & \{\caseeff{l}{x}{k}{e}\} \uplus h \\
    \text{Values} & v & \coloneqq & n \mid k \mid \clos{x}{e}{\env} \mid \effval{l}{k} \mid \exnval{l} \\
    \text{Frames} & r & \coloneqq & \farg{e}{\env} \mid \ffun{v} \mid \faritha{\odot}{e}{\env} \mid \farithb{\odot}{ℕ} \\
    \text{Environments} & \env & \coloneqq & \emptyset \mid \envext{\env}{x}{v} \\
  \end{array}
  \end{smathpar}
  \end{minipage}
  \begin{minipage}[t]{0.49\linewidth}
  \begin{smathpar}
  \begin{array}{rrcl}
    \text{Handler Closures} & \hc & \coloneqq & (h,\env) \\
    \text{Frame List} & \fl & \coloneqq & [] \mid r :: \fl \\
    \text{Fibers} & \fiber & \coloneqq & (\fl, \hc) \\
    \text{Continuations} & k & \coloneqq & [] \mid \fiber \kcons k \\
    \text{C stacks} & \cstack & \coloneqq & \cstacka{\fl}{\ostack}\\
    \text{OCaml stacks} & \ostack & \coloneqq & \ostacka{k}{\cstack} \mid \ostackemp \\
    \text{Stacks} & \stack & \coloneqq & \cstack \mid \ostack \\
    \text{Terms} & \term & \coloneqq & e \mid v \\
    \text{Configurations} & \config & \coloneqq & \configa{\tau}{\env}{\stack}
  \end{array}
  \end{smathpar}
	\end{minipage}
  \subcaption{Syntax of expressions and configurations}
  \label{sem:syntax}
\end{minipage}
\begin{minipage}{0.25\linewidth}
	\begin{smathpar}
		\begin{array}{c}
			\textsc{StepC} ~
			\inferrule{(\term, \env, \fl, \ostack) \cstep \config}
								{\configa{\term}{\env}{\cstacka{\fl}{\ostack}} \step \config} \\ \\
			\textsc{StepO} ~
			\inferrule{(\term, \env, k, \cstack) \ostep \config}
								{\configa{\term}{\env}{\ostacka{k}{\cstack}} \step \config}
		\end{array}
	\end{smathpar}
	\subcaption{Top-level reductions}
	\label{sem:toplevel}
\end{minipage}
\begin{minipage}{0.69\linewidth}
  \begin{smathpar}
    \begin{array}{rrcl}
      \textsc{Var}     & (x,\env, \fl) & \rightsquigarrow
                       & (\env(x), \env, \fl) \\
      \textsc{Arith1}  & (e_1 \odot e_2, \env, \fl) & \rightsquigarrow
                       & (e_1,\env, \faritha{\odot}{e_2}{\env}::\fl) \\
      \textsc{Arith2}  & (n_1, \_, \faritha{\odot}{e_2}{\env}::\fl) & \rightsquigarrow
                       & (e_2, \env, \farithb{\odot}{n_1}::\fl) \\
      \textsc{Arith3}  & (n_2, \env, \farithb{\odot}{n_1}::\fl) & \rightsquigarrow
                       & (\llbracket n_1 \odot n_2 \rrbracket, \env, \fl) \\
      \textsc{App1}    & (e_1 ~e_2, \env, \fl) & \rightsquigarrow
                       & (e_1, \env, \farg{e_2}{\env}::\fl) \\
      \textsc{App2}    & (\lam{x}{e}, \env, \fl) & \rightsquigarrow
                       & (\clos{x}{e}{\env}, \env, \fl) \\
      \textsc{App3}    & (\clos{x}{e_1}{\env_1}, \_, \farg{e_2}{\env_2}::\fl) & \rightsquigarrow
                       & (e_2, \env_2, \ffun{\clos{x}{e_1}{\env_1}}::\fl) \\
      \textsc{Resume1} & (k,\_,\farg{e_1}{\env_1}::\farg{e_2}{\env_2}::\fl) & \rightsquigarrow
                       & (e_1, \env_1, \ffun{k}::\farg{e_2}{\env_2}::\fl) \\
      \textsc{Resume2} & (\clos{x}{e_1}{\env_1}, \_, \ffun{k}::\farg{e_2}{\env_2}::\fl) & \rightsquigarrow
                       & (e_2, \env_2, \ffun{k}::\ffun{\clos{x}{e_1}{\env_1}}::\fl) \\
      \textsc{Perform} & (\perform{l}{e}, \env, \fl) & \rightsquigarrow
                       & (e, \env, \ffun{\effval{l}{[[],(\{\caseval{x}{x}\},\emptyset)]}}::\fl) \\
      \textsc{Raise}   & (\throw{l}{e}, \env, \fl) & \rightsquigarrow
                       & (e, \env, \ffun{\exnval{l}}::\fl)
    \end{array}
  \end{smathpar}
  \subcaption{Administrative Reductions -- $(\term, \env, \fl) \rightsquigarrow (\tau, \env, \fl)$.}
  \label{sem:step}
\end{minipage}
\begin{minipage}{\linewidth}
  \begin{smathpar}
    \begin{array}{rrcll}
      \textsc{AdminC}   & (\term,\env,\fl,\ostack) & \cstep
                        & \configa{\term'}{\env'}{\cstacka{\fl'}{\ostack}}
                        & \text{if } (\term,\env,\fl) \rightsquigarrow (\term',\env',\fl') \\
			\textsc{CallC}    & (v, \_, \ffun{\cclos{x}{e}{\env}}::\fl, \ostack) & \cstep
                        & \configa{e}{\envext{\env}{x}{v}}{\cstacka{\fl}{\ostack}} \\
			\textsc{Callback} & (v, \_, \ffun{\oclos{x}{e}{\env}}::\fl, \ostack) & \cstep
                        & \configa{e}{\envext{\env}{x}{v}}{\ostacka{k}{\cstacka{\fl}{\ostack}}}
                        & \text{if } k = [[],(\{\caseval{x}{x}\},\emptyset)] \\
      \textsc{RetToO}   & (v,\env,[],\ostacka{k}{\cstack}) & \cstep
                        & \configa{v}{\env}{\ostacka{k}{\cstack}} \\
      \textsc{ExnFwdO}  & (v, \env, \ffun{\exnval{l}}::\_, \ostacka{(\fl,\hc) \kcons k}{\cstack}) & \cstep
                        & \configa{v}{\env}{\ostacka{(\ffun{\exnval{l}}::\fl,\hc) \kcons k}{\cstack}}
    \end{array}
  \end{smathpar}
  \subcaption{C Reductions -- $(\term, \env, \fl, \ostack) \cstep \config$.}
  \label{sem:cstep}
\end{minipage}
\begin{minipage}{\linewidth}
  \begin{smathpar}
    \begin{array}{rrcll}
      \textsc{AdminO}   & (\term,\env,(\fl,\hc) \kcons k,\cstack) & \ostep
                        & \configa{\term'}{\env'}{\ostacka{(\fl',\hc) \kcons k}{\cstack}}
                        & \text{if } (\term,\env,\fl) \rightsquigarrow (\term',\env',\fl') \\
			\textsc{CallO}    & (v, \_, (\ffun{\oclos{x}{e}{\env}}::\fl,\hc) \kcons k, \cstack) & \ostep
                        & \configa{e}{\envext{\env}{x}{v}}{\ostacka{(\fl,\hc) \kcons k}{\cstack}} \\
			\textsc{ExtCall}  & (v, \_, (\ffun{\cclos{x}{e}{\env}}::\fl,\hc) \kcons k, \cstack) & \ostep
                        & \configa{e}{\envext{\env}{x}{v}}{\cstacka{[]}{\ostacka{(\fl,\hc) \kcons k}{\cstack}}} \\
      \textsc{RetToC}   & (v, \_, [([],(h,\emptyset))], \cstack) & \ostep
                        & \configa{v}{\env}{\cstack}
                        & \text{if } h = \{\caseval{x}{x}\} \\
      \textsc{RetFib} & (v, \_, ([],(h,\env)) \kcons k, \cstack) & \ostep
                        & \configa{e}{\envext{\env}{x}{v}}{\ostacka{k}{\cstack}}
                        & \text{if } \{\caseval{x}{e}\} \in h \\
									& & & & \text{and } k \neq [] \\

      \textsc{Handle}   & (\handle{e}{h}, \env, k, \cstack) & \ostep
                        & \configa{e}{\env}{\ostacka{([],(h,\env)) \kcons k}{\cstack}} \\
      \textsc{ExnHn}     & (v, \_, (\ffun{\exnval{l}}::\_,(h,\env)) \kcons k', \cstack) & \ostep
                        & \configa{e}{\envext{\env}{x}{v}}{\ostacka{k'}{\cstack}}
                        & \text{if } \{\caseexn{l}{x}{e}\} \in h \\
      \textsc{ExnFwdC}   & (v, \env, [\ffun{\exnval{l}}::\_,(h,\_)], \cstacka{\fl'}{\ostack}) & \ostep
                        & \configa{v}{\env}{\cstacka{\ffun{\exnval{l}}::\fl'}{\ostack}}
                        & \text{if } \{\caseexn{l}{\_}{\_}\} \notin h \\
      \textsc{ExnFwdFib} & (v, \env, (\ffun{\exnval{l}}::\_,(h,\_)) \kcons (\fl',\hc') \kcons k', \cstack) & \ostep
                        & \configa{v}{\env}{\ostacka{(\ffun{\exnval{l}}::\fl',\hc') \kcons k'}{\cstack}}
                        & \text{if } \{\caseexn{l}{\_}{\_}\} \notin h \\

      \textsc{EffHn}    & (v, \_, (\ffun{\effval{l}{k}}::\fl,(h,\env)) \kcons k', \cstack) & \ostep
                        & \configa{e}{\env[r \mapsto k''][x \mapsto v]}{\ostacka{k'}{\cstack}}
                        & \text{if } \{\caseeff{l}{x}{r}{e}\} \in h \\
                  & & & & \text{and } k'' = k ~@~ [(\fl,(h,\env))] \\
      \textsc{EffFwd}   & (v, \env', (\ffun{\effval{l}{k}}::\fl,(h,\env)) \kcons (\fl',\hc') \kcons k', \cstack) & \ostep
                        & \configa{v}{\env'}{\ostacka{(\ffun{\effval{l}{k''}}::\fl',\hc') \kcons k'}{\cstack}}
                        & \text{if } \{\caseeff{l}{\_}{\_}{\_}\} \notin h \\
                  & & & & \text{and } k'' = k ~@~ [(\fl,(h,\env))] \\

      \textsc{EffUnHn}  & (v, \_, [\ffun{\effval{l}{k}}::\fl,(h,\env)], \cstack) & \ostep
                        & \configa{e}{\emptyset}{\ostacka{k ~@~ [(\fl,(h,\env))]}{\cstack}}
                        & \text{if } \{\caseeff{l}{\_}{\_}{\_}\} \notin h \\
                  & & & & \text{and } {e = \throw{\textsf{Unhandled}}{0}} \\
      \textsc{Resume}   & (v, \_, (\ffun{k}::\ffun{\oclos{x}{e}{\env}}::\fl,\hc) \kcons k', \cstack) & \ostep
                        & \configa{e}{\envext{\env}{x}{v}}{\ostacka{k ~@~ ((\fl,\hc) \kcons k')}{\cstack}} \\
    \end{array}
  \end{smathpar}
  \subcaption{OCaml Reductions -- $(\term, \env, k, \cstack) \ostep \config$.}
  \label{sem:ostep}
\end{minipage}
\caption{Operational semantics of Multicore OCaml effect handlers.}
\end{figure*}

\vspace{-1mm}
\subsection{Dynamic semantics}

We present an operational semantics for a core language of effect handlers that
faithfully captures the semantics of the Multicore OCaml implementation. An
executable version of the semantics, implemented as an OCaml interpreter, along
with examples, is included in the supplementary material.

\vspace{-1mm}
\subsubsection{Syntax}

Our expressions (Figure~\ref{sem:syntax}) consist of integer constants ($n$),
variables ($x$), abstraction ($\lam{x}{e}$), application ($e~e$), arithmetic
expressions ($e \odot e$) where $\odot$ ranges over $\{+,-,*,/\}$, raising
exceptions ($\throw{l}{e}$), performing effects ($\perform{l}{e}$), and
handling effects ($\handle{e}{h}$). Abstractions come in two forms: OCaml
abstractions ($\lambda^o$) and C abstractions ($\lambda^c$). The handler
consists of a return case ($\caseval{x}{e}$), zero or more exception cases
($\caseexn{l}{x}{e}$) with label $l$, parameter $x$ and body $e$, and zero or
more effect cases ($\caseeff{l}{x}{k}{e}$) with label $l$, parameter $x$,
continuation $k$ and body $e$.

The operational semantics is an extension of the CEK machine
semantics~\cite{Felleisen86} for effect handlers, following the abstract
machine semantics of Hillerstrom et al.~\cite{Hillerstrom20}. The key
difference from Hillerstrom et al. is that our stacks are composed of
alternating sequence of OCaml and C stack segments. The program state is
captured as configuration $\config \coloneqq \configa{\tau}{\env}{\stack}$ with
the current term $\tau$ under evaluation, its environment $\env$ and the
current stack $\stack$. The term is either an expression $e$ or a value $v$.
The values are integer constants $n$, continuations $k$, closures
$\clos{x}{e}{\env}$, effects being performed ($\effval{l}{k}$) and exceptions
being raised ($\exnval{l}$). The environment is a map from variables to values.

The stack $\stack$ is either a C stack ($\cstack$) or an OCaml stack
($\ostack$). The C stack $\cstacka{\fl}{\ostack}$ consists of a list of
frames $\fl$, and the OCaml stack $\ostack$ under it. The OCaml stack is
either empty $\ostackemp$ or non-empty $\ostacka{k}{\cstack}$ with the current
continuation $k$ and the C stack $\cstack$ under it. Thus, the program stack is
an alternating sequence of C and OCaml stacks terminating with an empty OCaml
stack $\ostackemp$. The frame list $\fl$ is composed of individual frames $r$,
which is one of an argument frame $\farg{e}{\env}$ with the expression $e$ at
the argument position of an application with its environment $\env$, a function
frame $\ffun{v}$ with the value $v$ at the function position of an application,
and frames for evaluating the arguments of an arithmetic expression.

A continuation $k$ is either empty $[]$ or a non-empty list of \emph{fibers}. A
fiber $\fiber \coloneqq (\fl,\hc)$ is a list of frames $\fl$ and a handler
closure $\hc \coloneqq (h,\env)$, which is a pair of handler $h$ and its
environment $\env$. We use the infix operator $@$ for appending two lists.

\vspace{-1mm}
\subsubsection{Top-level reductions}

The initial configuration for an expression $e$ is
$\configa{e}{\emptyset}{\cstacka{[]}{\bullet}}$, where the environment and the
stack are empty. The top-level reductions (Figure ~\ref{sem:toplevel}) can be
performed by either by taking a C step $\cstep$ or an OCaml step $\ostep$.

\vspace{-1mm}
\subsubsection{C reductions}

We can take a C step (Figure~\ref{sem:cstep}) by taking an administrative
reduction step $\rightsquigarrow$. The administrative reductions are common to
both C and OCaml. The rules \textsc{Var}, \textsc{Arith1}, \textsc{Arith2},
\textsc{App1}, \textsc{App2} and \textsc{App3} are standard. \textsc{Arith3}
performs the arithmetic operation on the integers ($\llbracket n_1 \odot n_2
\rrbracket$). \textsc{Raise} pushes an function frame with exception value to
indicate that an exception is being raised. Similarly, \textsc{Perform} pushes
a function frame with an effect value with an empty continuation
$[[],(\{\caseval{x}{x}\},\emptyset)]$ with no captured frames and an empty
handler with an identity return case alone. We shall return to \textsc{Resume1}
and \textsc{Resume2} in the next subsection.

Continuing with the rest of the C reduction steps, \textsc{CallC} captures the
behaviour of calling a C function. Since the program is currently executing C,
we can perform the call on the current stack. In case the abstraction is an
OCaml abstraction (\textsc{Callback}), we create an OCaml stack with the C
stack as its tail, with the current continuation being empty. This captures the
behaviour of calling back into OCaml from C. \textsc{RetToO} returns a value to
the enclosing OCaml stack. \textsc{ExnFwdO} forwards a raised exception to the
enclosing OCaml stack, unwinding the rest of the frames. This captures the
semantics of raising OCaml exceptions from C.

\vspace{-1mm}
\subsubsection{OCaml reductions}

In OCaml (Figure~\ref{sem:ostep}), reductions always occur on the top-most
fiber in the current stack. \textsc{AdminO} performs administrative reductions.
\textsc{CallO} evaluates an OCaml function on the current stack.
\textsc{ExtCall} captures the behaviour of external calls, which are evaluated
on an empty C stack with the current OCaml stack as its tail. \textsc{RetToC}
returns a value to the enclosing C stack. In this case, we have exactly one
fiber on the stack, and this was created in the rule \textsc{Callback}, whose
handler has identity return case alone and the environment is empty.
\textsc{RetFib} returns the value from a fiber to the previous one, evaluating the
body of the return case.

The rule \textsc{Handle} installs a handler by pushing a fiber with no frames
and the given handler. The rule \textsc{ExnHn} handles an exception, if the
current handler has a matching exception case, unwinding the current fiber. The
rule \textsc{ExnFwdC} forwards the exception to C. Here, there is exactly one
fiber on the current stack, and the handler does not have a matching exception
case, which we know is the case (see \textsc{Callback} rule). The rule
\textsc{ExnFwdFib} forwards the exception to the next fiber if the current
handler does not handle it.

The rule \textsc{EffHn} captures the handling of effects when the current
handler has a matching effect case. We evaluate the body of the matching case,
and bind the continuation parameter |r| to the captured continuation |k''|.
Observe that the captured continuation |k''| includes the current handler.
Intuitively, the handler wraps around captured continuation. This gives
Multicore OCaml effect handlers deep handler semantics~\cite{Hillerstrom20}.
\textsc{EffFwd} forwards the effect to the outer fiber, and extends the
captured continuation |k''| in the process. Recall that we do not capture C
frames as part of a continuation. To this end, \textsc{EffUnHn} models
unhandled effect. If the effect bubbles up to the top fiber --- which we know does
not have an effect case (see \textsc{Callback} rule) --- we raise
\textsc{Unhandled} exception at the point where the corresponding effect was
performed. This is achieved by appending the captured continuation to the front
of the current continuation.

Observe that |continue| and |discontinue| are not part of the expressions. They
are encoded as $\textsf{continue} ~k ~e = (k ~(\olam{x}{x})) ~e$ and
$\textsf{discontinue} ~k ~l ~e = (k ~(\olam{x}{\throw{l}{x}})) ~e$.
Intuitively, resuming a continuation in both the cases involves evaluating the
appropriate abstraction on top of the continuation. We perform the
administrative reductions \textsc{Resume1} and \textsc{Resume2} to evaluate the
arguments to |continue| and |discontinue|. The rule \textsc{Resume} appends the
given continuation to the front of the current continuation, and evaluates the
body of the closure.

\vspace{-2mm}
\section{Implementation}
\label{sec:impl}

We now present the implementation details of effect handlers in Multicore
OCaml. While we assume an x86\_64 architecture for the remainder of this paper,
our design does not preclude other architectures and operating systems.

\subsection{Exceptions}

The implementation follows the operational semantics, but has a few key
representational differences. Unlike the operational semantics, handlers with
just exception patterns (exception handlers) are implemented differently than
effect handlers. As mentioned in \S\ref{sec:exn_handlers}, exceptions are
pervasive in OCaml and are so cheap that they are used for local control flow.
Hence, we retain the linked exception handler frame implementation of stock
OCaml in Multicore OCaml to ensure performance backwards compatibility. This
differs from other research languages with effect
handlers~\cite{Hillerstrom20,Frank,Eff}, which implement exceptions using
effects (by ignoring the continuation argument in the handler).

\subsection{Heap-allocated fibers}
\label{sec:hafibers}

\begin{figure*}
\begin{minipage}{0.35\linewidth}
  \centering
  \includegraphics[scale=0.45]{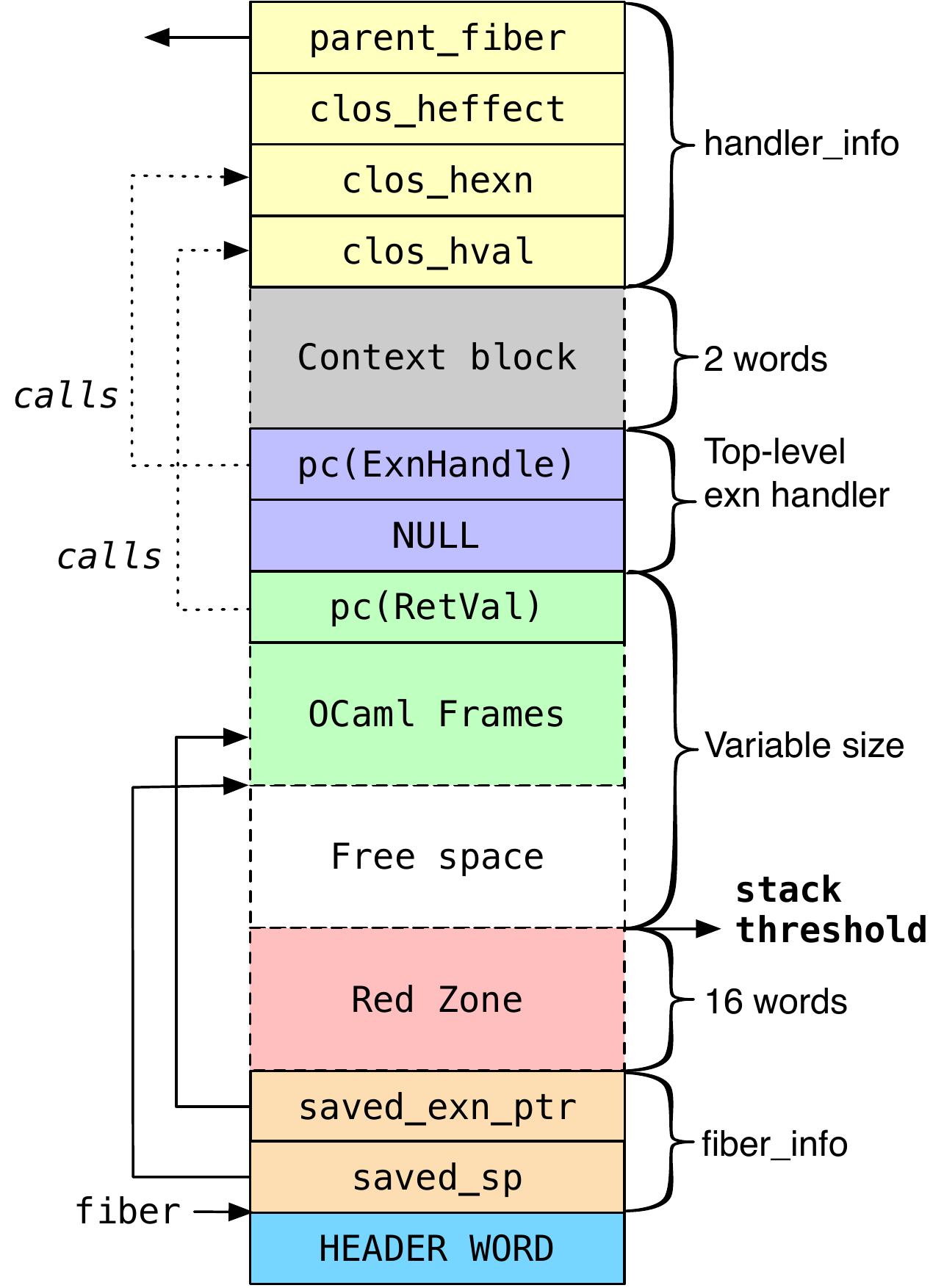}
  \subcaption{Fiber layout}
  \label{fig:fiber}
\end{minipage}
\begin{minipage}{0.64\linewidth}
  \begin{minipage}{\linewidth}
    \centering
    \includegraphics[scale=0.4]{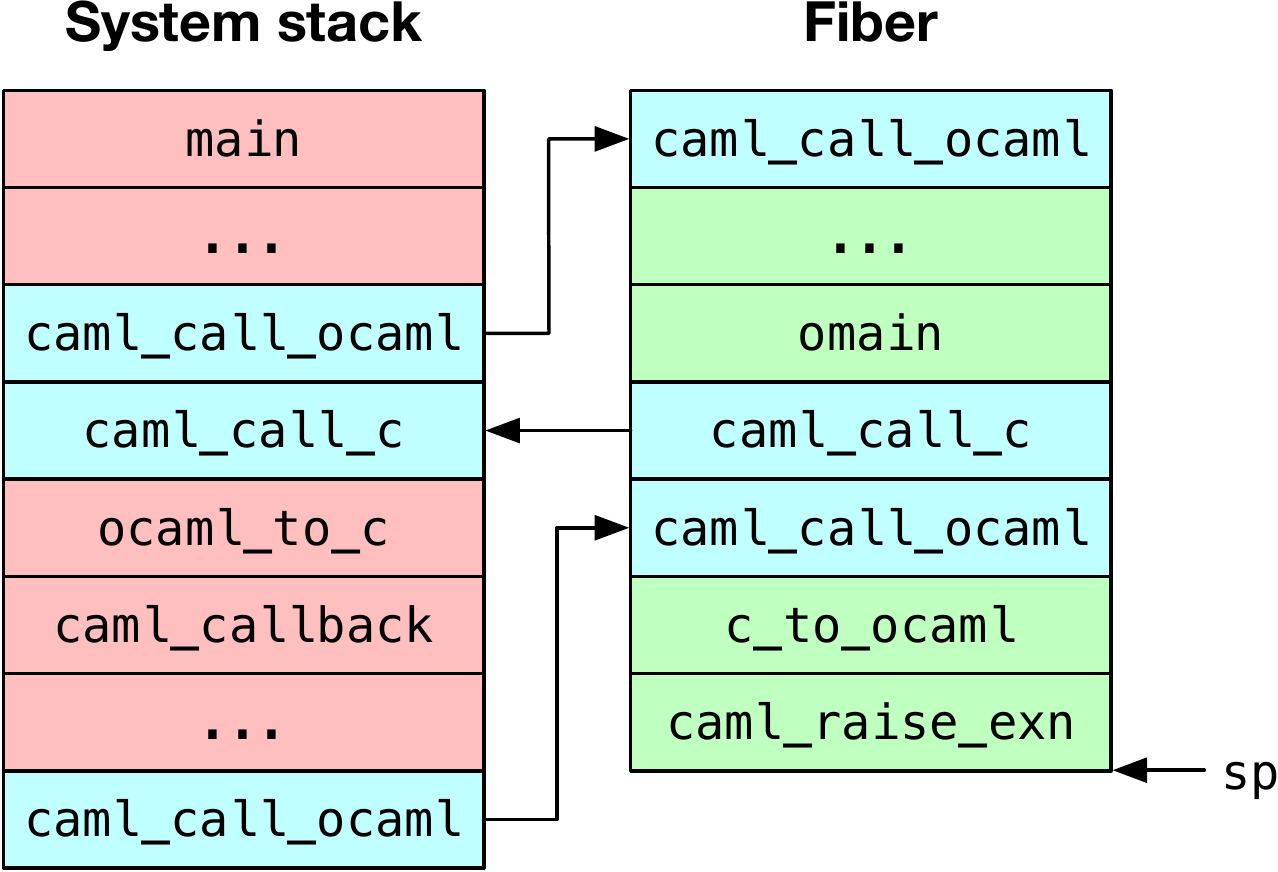}
    \subcaption{Stack layout for \texttt{meander} example from \S\ref{sec:stack}.}
    \label{fig:mcstack}
  \end{minipage}
  \begin{minipage}{0.55\linewidth}
		\vspace{1mm}
    \begin{lstlisting}
effect E : unit
effect F : unit;;

match (* comp_e *)
  match (* comp_f *)
    (*p1*) perform E (*p3*)
	with | v -> v | effect F kf -> ()
with | v -> v
		 | effect E ke -> (*p2*) continue ke ()
    \end{lstlisting}
		\subcaption{Constructing continuation objects}
		\label{code:effimpl}
  \end{minipage}
  \begin{minipage}{0.44\linewidth}
		\vspace{2mm}
    \centering
    \includegraphics[scale=0.42]{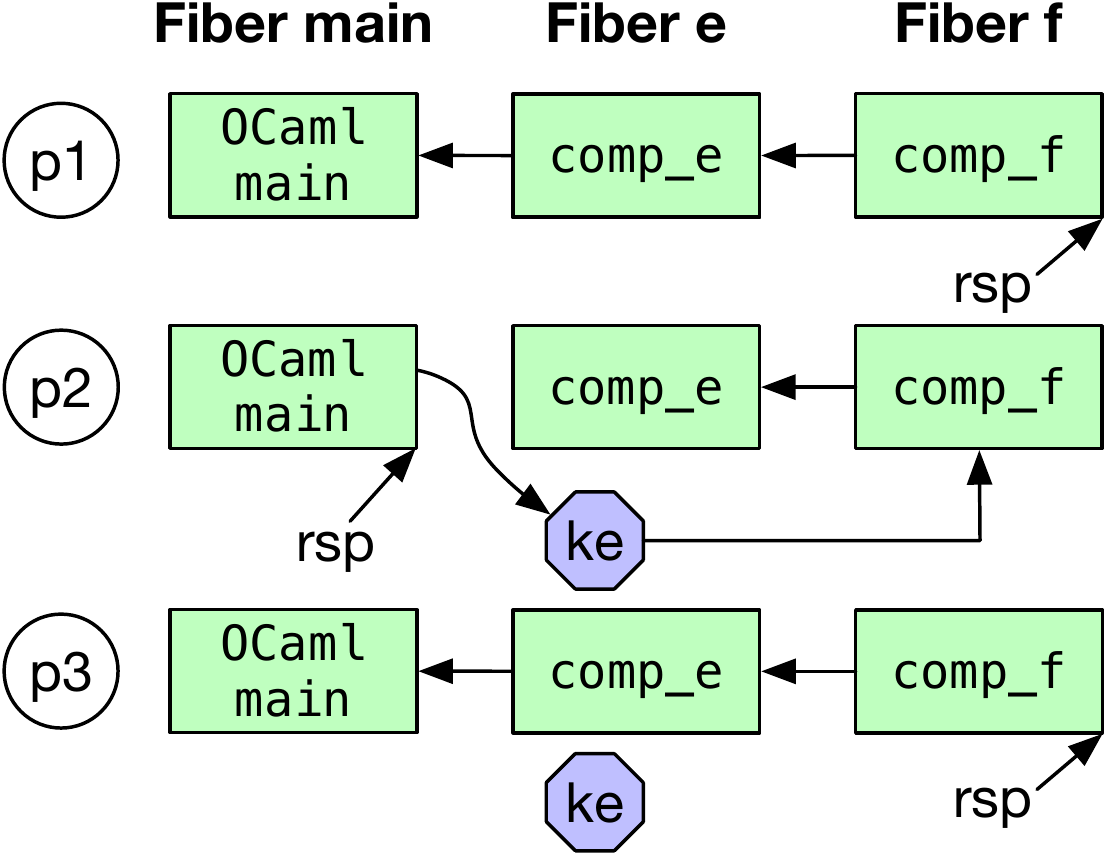}
		\subcaption{Program state for code in~\ref{code:effimpl}}
    \label{fig:fiber_handler}
  \end{minipage}
\end{minipage}
\vspace{-2mm}
\caption{Layout of Multicore OCaml effect handlers.}
\vspace{-3mm}
\end{figure*}

In the operational semantics, the continuations may be resumed more than once.
Captured continuations are copies of the original fibers and resuming the
continuation copies the fibers and leaves the continuation as it is. Since our
primary use case is concurrency, continuations will be resumed at most once, and
copying fibers is unnecessary and inefficient. Instead, Multicore OCaml
optimises fibers for one-shot continuations. Fibers are allocated on the C heap
using |malloc| and are |free|d when the handled computation returns with a value
or an exception. Similar to Farvardin et al.~\cite{Farvardin20}, we use a
\emph{stack cache} of recently freed stacks in order to speed up allocation.

Figure~\ref{fig:fiber} shows the layout of a fiber in Multicore OCaml. At the
bottom of the stack, we have the |handler_info|, which contains the pointer to
the parent fiber, and the closures for the value, exception and effect cases.
The closures are created by the translation described in
\S\ref{sec:static_semantics}; Multicore OCaml supports exception patterns in
addition to effect patterns in the same handler. This is followed by a context
block needed for DWARF and GC bookkeeping with callbacks. Then, there is a
top-level exception handler frame that forwards exceptions to the parent fiber.
When the exceptions are caught by this handler, the control switches to the
parent stack, and the exception handler closure |clos_hexn| is invoked. This is
followed by the |pc| of the code that returns values to the parent
fiber. This stack is laid out such that when the handled computation returns,
the control switches to the parent fiber and the value handler |clos_hval|
is invoked.

Next, we have the variable-sized area for the OCaml frames. In order to keep
fibers small, this area is initially 16 words in length. When the stack pointer
|rsp| becomes less than the \emph{stack threshold} (maintained in the
|Caml_state| table), the stack is said to have overflowed. On stack overflow,
we copy the whole fiber to a new area with double the size. In Multicore OCaml,
we introduce stack overflow checks into the function prologue of OCaml
functions. These stack overflows are rare and so the overflow checks will be
correctly predicted by the CPU branch predictor.

In our evaluation of real world OCaml programs (\S\ref{sec:eval}), we observed
that most function calls are to leaf functions with small frame sizes. Can we
eliminate the stack overflow checks for these functions? To this end, we
introduce a small, fixed-sized red zone at the top of the stack. The compiler
elides the stack overflow check for leaf functions whose frame size is less
than the size of the red zone. The default size of the red zone in Multicore
OCaml is 16 words.

Finally, we have the saved exception pointer, which points to the top-most
exception frame, and the saved stack pointer, which points to the top of the
stack. Switching between fibers only involves saving the exception and the
stack pointer of the current stack and loading the same on the target stack.
Since OCaml does not generate pointers into the stack, the two |fiber_info|
fields are the only ones that need to be updated when fibers are moved.

\subsection{External calls and callbacks}

Since C functions do not have stack overflow checks, we have to execute the
external calls in the system stack. Calling a C function from OCaml involves
saving the stack pointer in the current fiber, saving the allocation pointer
value in |r15| in the |Caml_state|, updating |rsp| to the top of system stack
(maintained in |Caml_state|), and calling the C function. The actions are
reversed when returning from the external call. For C functions that take
arguments on the stack, the arguments must be copied to the C stack from the
OCaml stack.

When we first enter OCaml from C, a new fiber is allocated for the main OCaml
stack. Since callbacks may be frequent in OCaml programs that use finalisers,
we run the callbacks on the same fiber as the current one. For example, the
layout of the Multicore OCaml stack at |caml_raise_exn| in the |meander|
example from \S\ref{sec:stack} is shown in Figure~\ref{fig:mcstack}. The
functions |caml_call_c| and |caml_call_ocaml| switch the stacks, and hence are
shown in both the system stack and the fiber. Since we are reusing the fiber
for the callback, care must be taken to save and restore the |handler_info|
before calling and after returning from |c_to_ocaml| function, respectively.
Thanks to the fiber representation, external calls and callbacks remain
competitive with stock OCaml.

\subsection{Effect handlers}
\label{sec:effimpl}

Similar to exception handlers, the lack of callee-saved registers in OCaml
benefits effect handlers. There is no register state to save when entering an
effect handler or performing an effect. Similarly, there is no register state
to restore when handling an effect or resuming a continuation. This fortuitous
design choice in stock OCaml has a significant impact in enabling fast
switching between fibers in Multicore OCaml.

In order to illustrate the runtime support for handling effects, consider the
example presented in Figure~\ref{code:effimpl}. The layout of the program state
as the program executes is captured in Figure~\ref{fig:fiber_handler}. The code
performs effect |E| which is handled in the outer-most handler, and is
immediately resumed. The arrows between the fibers are parent pointers. At
position |p1|, |rsp| is at the top of the fiber |f|.

When the effect |E| is performed, we allocate a continuation object |ke| in the
OCaml heap that points to the current fiber |f|, set fiber |f|'s parent pointer
to |NULL|, and evaluate the continuation closure |clos_heffect| on the parent
fiber |e| with the effect |E| and the continuation |ke| as arguments. Since the
first handler does not handle effect |E|, the effect is \emph{reperformed}
(\S\ref{sec:static_semantics}) by appending the fiber |e| to the tail of
continuation |ke|, set fiber |e|'s parent pointer to |NULL|, and evaluate the
current continuation closure on the parent fiber |main| with |E| and |ke| as
arguments, which handles |E| (position |p2|). Thus, \emph{continuations are
captured without copying frames}. Since every handler closure is evaluated
until a matching one is found, the time taken to handle an effect is linear in
the number of handlers. We observed that the handler stack is shallow in real
programs.

When the continuation is resumed, we overwrite the value of |ke| to |NULL| to
enforce at-most once semantics. Resuming a continuation involves traversing the
linked-list of fibers and making the last fiber point to the current fiber.
Just as in the operational semantics, the implementation invokes the
appropriate closure to either |continue| or |discontinue| the continuation
(position |p3|). We perform tail-call optimisation so that resumptions at tail
positions do not build up stack.

\vspace{-2mm}
\subsection{Stack unwinding}

The challenge with DWARF stack unwinding is to make it aware of the
non-contiguous stacks. While the complete details of DWARF stack unwinding is
beyond the scope of the paper, it is beneficial to know how DWARF unwind tables
are constructed in order to appreciate our solution. We refer the interested
reader to Bastian et al.~\cite{Bastian19} for a good overview of DWARF stack
unwinding.

Logically, DWARF call-frame information maintains a large table which records
for every machine instruction where the return address and callee-saved
registers are stored. To avoid reifying this large table, DWARF directives
represent the table using a compact bytecode representation that describes the
unwind table as a sequence of edits from the start of the function. In order to
compute the call-frame information at any given instruction within a function,
the DWARF bytecode from the start of that function must be interpreted on
demand. For each function, DWARF maintains a \emph{canonical frame address}
(CFA) and is traditionally the stack pointer before entering this function.
Hence, on x86-64, where the return address is pushed on the stack on call, the
return address is at |CFA - 8|.

Our goal is to compute the CFA of the caller when stacks are switched using the
DWARF directives. Recall that stack switching occurs in effect handlers,
external calls and callbacks. At the entry to an effect handler block, we insert
DWARF bytecode to follow the |parent_fiber| pointer and dereference the
|saved_sp| to get the CFA (|saved_sp + 8|). During callbacks into OCaml, we
save the current system stack pointer in the |context block| in
Figure~\ref{fig:fiber} to identify the CFA in the C stack. DWARF unwinding for
external calls is implemented by following a link to the current OCaml stack
pointer. With these changes, we get the same backtrace for the \texttt{meander}
program from \S\ref{sec:unwind}, modulo runtime system functions due to effect
handlers. We have verified the correctness of our DWARF directives using the
verification tool from Bastian et al.~\cite{Bastian19}.

Despite the correct DWARF unwind information, using DWARF to record call stack
information in |perf| only captures the call stack of the current fiber in
Multicore OCaml. Since stack unwinding using DWARF is slow due to bytecode
interpretation overhead, |perf| dumps the (user) call stack when
sampled~\cite{Bastian19}. This only includes the frames from the current fiber.
This is a limitation of |perf| and not of our stack layout. Bastian et
al.~\cite{Bastian19} report on a technique to pre-compile the unwind table to
assembly, which speeds up DWARF-based unwinding by up to 25$\times$. With this
technique, |perf| can unwind the stack at sample points rather than dumping the
call stack, which would capture the complete backtrace rather than just the
current fiber.

\vspace{-1mm}
\subsection{Garbage collection}
\label{sec:gc}

Recall that OCaml programs are written with the expectation that function calls
return exactly once (\S\ref{sec:refine}). Consider the scenario when a
continuation is never resumed. Since fibers allocate memory for the stack using
|malloc|, which are |free|d when the computation returns, not resuming
continuations leaks memory. In addition, unresumed continuations may also leak
other system resources such as sockets and open file descriptors.

We make a pragmatic trade-off and expect the user code to resume captured
continuations \emph{exactly once}. One can use the GC support to free up
resources by installing a finaliser that discontinues the continuation and
ignores the result:

\begin{lstlisting}
Gc.finalise (fun k ->
  try ignore (discontinue k Unwind) with _ -> ()) k
\end{lstlisting}

This frees up both the memory allocated for the fiber stack as well as other
system resources, assuming that user code does not handle |Unwind| exception
and fails to re-raise it. Since installing a finaliser on every captured
continuation introduces significant overhead (\S\ref{sec:res_final_k}), we
choose not to do it by default. It is also useful to note that even if the
memory for the fiber stack is managed by the GC, we would still need a
finalisation mechanism to unwind the stack and release other system resources
that may be held by the continuation.

The challenges and the solutions for integrating fibers with the concurrent
mark-and-sweep GC of Multicore OCaml have been discussed
previously~\cite{Sivaramakrishnan20}.

\section{Evaluation}
\label{sec:eval}

In this section, we evaluate the performance of Multicore OCaml effect handlers
against the performance requirements set in \S\ref{sec:req}. Multicore OCaml is
an extension of the OCaml 4.10.0 compiler with support for shared memory
parallelism and effect handlers. Since our objective is to evaluate the impact
of effect handlers, none of our benchmarks utilise parallelism. These results
were obtained on a 2-socket Intel\textregistered Xeon\textregistered Gold 5120
x86-64~\cite{IntelXeonGold5120Spec} server running Ubuntu 18.04 with 64GB of
main memory.

\vspace{-1mm}
\subsection{No effects benchmarks}

In this section, we measure the impact of the addition of effect handlers on
code that does not use effect handlers. Our macro benchmark suite consists of
54 real OCaml workloads including verification tools (|Coq|, |Cubicle|,
|AltErgo|), parsers (|menhir|, |yojson|), storage engines (|irmin|), utilities
(|cpdf|, |decompress|), bioinformatics (|fasta|, |knucleotide|, |revcomp2|,
|regexredux2|), numerical analysis (|grammatrix|, |LU_decomposition|) and
simulations (|nbody|, |game_of_life|). In addition to Stock (|stock|) and
Multicore OCaml (|MC|), we also ran the benchmarks on Multicore OCaml with no
red zone (|MC+RedZone0|) in the fibers (all OCaml functions will have a stack
overflow check) and a red zone size of 32 words (|MC+RedZone32|). Recall that
the default red zone size in Multicore OCaml is 16 words (\S\ref{sec:hafibers}).

\begin{figure*}
	\includegraphics[scale=0.34]{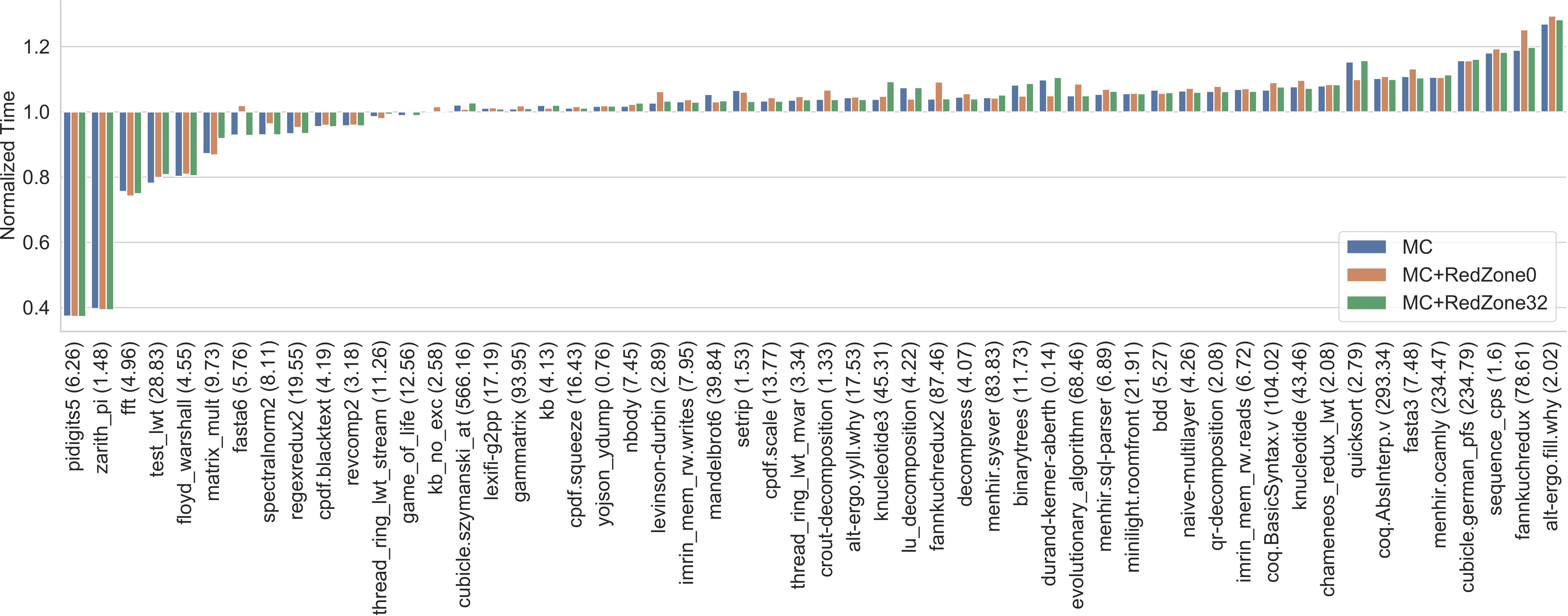}
	\vspace{-4mm}
	\caption{Normalized time of macro benchmarks. Baseline is Stock OCaml,
		whose running time in seconds in given in parenthesis.}
	\vspace{-2mm}
	\label{res:macro_time}
\end{figure*}

\begin{figure*}
	\includegraphics[scale=0.34]{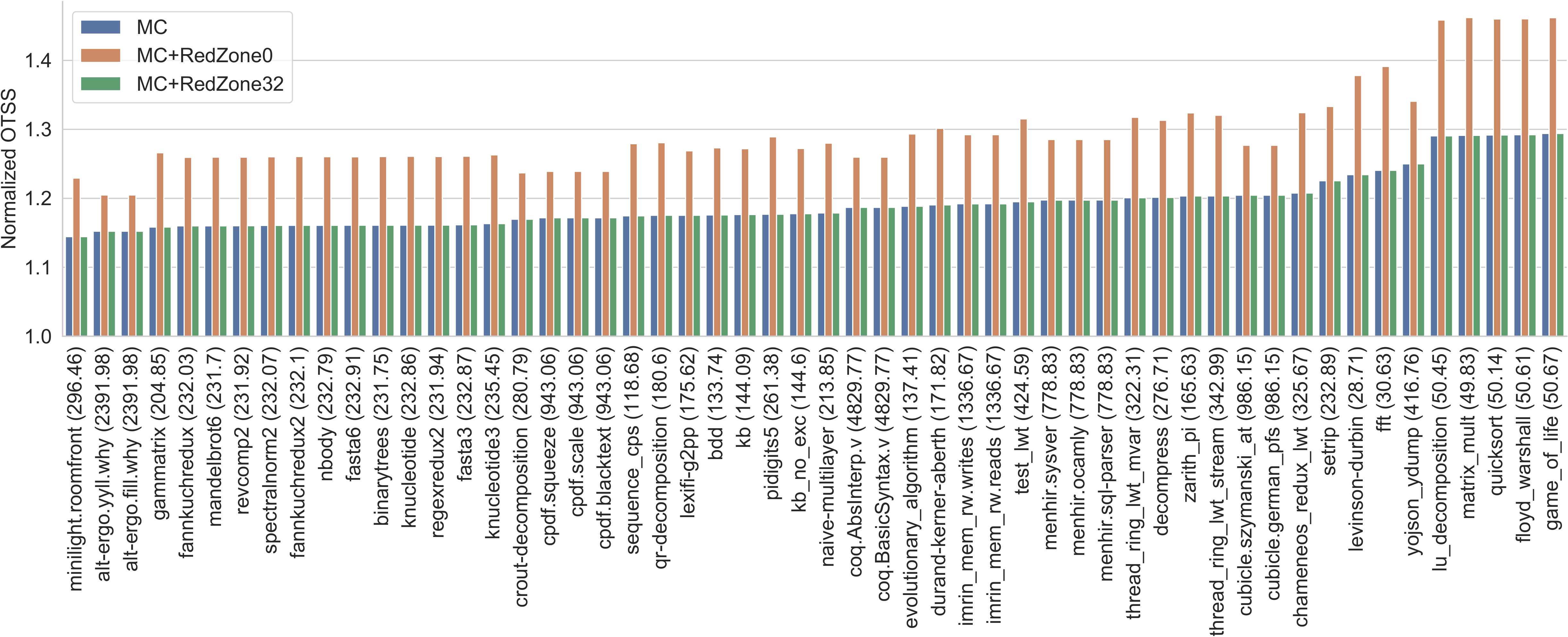}
	\vspace{-4mm}
	\caption{Normalized OCaml text section size (OTSS) of macro benchmarks.
	Baseline is Stock OCaml, whose size in kilobytes given in parenthesis.}
	\vspace{-2mm}
	\label{res:macro_size}
\end{figure*}

Figure~\ref{res:macro_time} presents the running time of the different
multicore variants normalized against the sequential baseline |stock|. On
average (geometric mean of the normalized values against |stock| as the
baseline), the multicore variants were less than 1\% slower than |stock|. The
outliers (on either ends) were due to the difference in the allocator and the
GC between stock and Multicore OCaml. Of the 54 programs in the benchmark
suite, 32 programs had an overhead of 5\% or lower, and 8 programs had more
than 10\% overhead.

The biggest impact was the increase in the OCaml text section size (OTSS) due
to the stack overflow checks. We define OTSS as the sum of the sizes of all the
OCaml text sections in the compiled binary file ignoring the data sections, the
debug symbols, the text sections associated with OCaml runtime and other
statically linked C libraries. Figure~\ref{res:macro_size} presents the OTSS of
the multicore variants normalized against the sequential baseline |stock|.
Compared to |stock|, OTSS is 19\% more for |MC| and |MC+RedZone32|, and 30\%
more for |MC+RedZone0|. The result shows that our 16-word red zone is effective
at reducing OTSS compared to having no red zone, whereas the 32-word red zone
does not noticeably reduce OTSS further. Further work is required to bring OTSS
closer to |stock|.

\begin{table}
\caption{Micro benchmarks without effects. Each entry is the percentage
	difference for Multicore OCaml over stock OCaml.}
\vspace{-5mm}
{
\begin{tabular}{r c c c c c c c c c c}
	& \rot{exnval} & \rot{exnraise} & \rot{extcall} & \rot{callback} & \rot{ack}
	& \rot{fib} & \rot{motzkin} & \rot{sudan} & \rot{tak} \\ \hline
	\textbf{Time} & +0.0 & -1.9 & +17 & +65  & +5.3
								& +2.2 & +10 & +0.0 & +4.2 \\
	\textbf{Instr} & +0.0 & +0.0 & +10 & +72 & +16
								 & +24 & +16 & +14 & +17 \\ \hline
\end{tabular}
}
\label{tab:micro_noeffect}
\vspace{-5mm}
\end{table}

We also present micro benchmarks results in Table~\ref{tab:micro_noeffect}.
Since micro benchmarks magnify micro-architectural optimisations, we also
report the number of instructions executed (obtained using |perf|) along with
time. |exnval| performs 100 million iterations of installing exception handlers
and returning with a value. |exnraise| is similar, but raises an exception in
each iteration. |extcall| and |callback| perform 100 million external calls and
callbacks to identity functions. The other micro benchmarks are highly
recursive programs and were taken from Farvardin et al.~\cite{Farvardin20}. For
micro benchmarks, we observed that padding tight loops with a few |nop|
instructions, which changes the loop alignment, makes the code up to 15\%
faster. Hence, the difference in running times under 15\% may not be
statistically significant.

The results show that exceptions are no more expensive in |MC| compared to
|stock|. In the other programs, |MC| executes more instructions due to stack
overflow checks. The performance impact on callbacks is more significant than
external calls. For callbacks, since we reuse the current fiber stack, we need
to ensure it has enough room for inserting additional frames, while |stock|
does not need to do this. Callback performance is less important than external
calls, which are far more numerous.

\vspace{-2mm}
\subsection{No perform benchmarks}

\begin{table}
\caption{Micro benchmarks with handlers but no perform. Each entry is the
	slowdown factor ($\times$ times) over its idiomatic implementation in stock OCaml.}
\vspace{-3mm}
{
\begin{tabular}{r c c c c c}
	& \textbf{ack} & \textbf{fib} & \textbf{motzkin} & \textbf{sudan} & \textbf{tak} \\ \hline
	\texttt{MC} 	 	& 12.25 & 12.05 & 11.44 & 6.74 & 8.9 \\
	\texttt{monad} 	& 348.69 & 69.77 & 39.24 & 33.29 & 42.79 \\ \hline
\end{tabular}
}
\label{tab:micro_noperform}
\vspace{-3mm}
\end{table}

Next, we aim to quantify the overhead of setting up and tearing down effect
handlers compared to a non-tail function call. To this end, we surround the
non-tail calls in the recursive micro benchmarks with an effect handler. These
programs do not |perform| effects. We also implemented the same benchmarks using
a concurrency monad~\cite{Claessen99} (|monad|) as a proxy for CPS versions.
Recall that the OCaml compiler does not use CPS in its IR. In the |monad|
version, we use a |fork| to invoke the non-tail call and use an |MVar| to
collect its result.

The results are presented in Table~\ref{tab:micro_noperform}. They show that
using effect handlers (concurrency monad) is 10.02$\times$ (67.09$\times$) more
expensive than the idiomatic implementation using non-tail calls. The
concurrency monad suffers due to the heap allocation of continuation frames
(which need to be garbage collected), whereas effect handlers benefit from
stack allocation of the frames. For example, the number of major collections
for the |ack| benchmark is 0 for stock OCaml, 1 for |MC| and 112 for |monad|.
Our concurrency monad (and other monadic concurrency libraries such as
Lwt~\cite{lwt} and Async~\cite[\S18]{rwo}) also have other downsides -- exceptions,
backtraces, and DWARF unwinding are no longer useful due to the lack of a
stack.

We note that a compiler that uses CPS IR will be faster than the concurrency
monad implementation due to optimisations to reduce the heap allocation of
continuation frames. But Farvardin et al.~\cite{Farvardin20} show that CPS with
optimisations is still slower than using the call stack.

\subsection{Concurrent benchmarks}

Next we look at benchmarks that utilize non-local control flow using effect
handlers. First, we quantify the cost of individual operations in effect
handling. Consider the following annotated code:
\begin{lstlisting}
effect E : unit
(* a *) match (* b *) perform E (* d *) with
| v -> (* e *) v
| effect E k -> (* c *) continue k ()
\end{lstlisting}
The sequence \texttt{a-b} involves allocating a new fiber and switching to it.
\texttt{b-c} is performing the effect and handling it. \texttt{c-d} is resuming
the continuation. \texttt{d-e} is returning from the fiber with a value and
freeing the fiber. We measured the time taken to execute these sequences using
|perf| support for cycle-accurate tracing on modern Intel processors. We
executed 10 iterations of the code, with 3 warm-up runs. For calibration, the
idle memory load latency for the local NUMA node is 93.2 ns as measured using
the Intel MLC tool~\cite{mlc}. We observed that the sequences \texttt{a-b},
\texttt{b-c}, \texttt{c-d} and \texttt{d-e} took \textbf{23 ns, 5 ns, 11 ns and
7 ns}, respectively. The time in the sequence \texttt{a-b} is dominated by the
memory allocation. Thus, the individual operations in effect handling are fast.

\subsubsection{Generators}

Generators allow data structures to be traversed on demand. Many languages
including JavaScript and Python provide generators as a built-in primitive.
Using effect handlers (|MC|), given any data structure (|'a t|) and its
iterator (|val iter: 'a t -> ('a -> unit) -> unit|), we can derive its
generator function (|val next : unit -> 'a option|)\footnote{https://gist.github.com/kayceesrk/eb0ab496c22861f21b1d9484772e982d}.
We evaluate the performance of traversing a complete binary tree of depth 25
using this generator. This involves $2^{26}$ stack switches in total. For
comparison, we implemented a hand-written, selective CPSed~\cite{Nielson01},
defunctionalised~\cite{Danvy01} version (|cps|) and a concurrency monad
(|monad|) version of the generator for the tree. Both |cps| and |monad|
versions are specialised to the binary tree with the usual caveats of not using
the stack for function calls. We observed that the |cps| version was the
fastest, thanks to specialisation and hand optimisation. |MC| version was only
2.76$\times$ slower than |cps| while being a generic solution, and the |monad|
version was 8.69$\times$ slower than |cps|.

\subsubsection{Chameneos}

Chameneos~\cite{Chameneos} is a concurrency game aimed at measuring context
switching and synchronization overheads. Our implementation uses |MVars| for
synchronization. We compare effect handler (|MC|), concurrency monad (|monad|)
and Lwt, a widely used concurrency programming library for OCaml (|lwt|)
versions. We observed that |MC| was the fastest, and |monad| (|lwt|) was
1.67$\times$ (4.29$\times$) slower than |MC|.

\subsubsection{Finalised continuations}
\label{sec:res_final_k}

In \S\ref{sec:gc}, we described how continuation resources can be cleaned up by
attaching a finaliser. Attaching this finaliser to every captured continuation
slows down generator (chameneos) benchmark by 4.1$\times$ (2.1$\times$)
compared to not attaching a finaliser. Hence, Multicore OCaml does not attach
such finaliser to every continuation by default.

\subsubsection{Webserver}

Using effect handlers, we have implemented a full-fledged HTTP/1.1 web server
by extending the example from \S\ref{sec:async_io} (|MC|). The web server
spawns a lightweight thread per request. We use \texttt{httpaf}~\cite{httpaf}
for HTTP handling, and \texttt{libev}~\cite{libev} for the eventloop. We
compare our implementation against an Lwt version (|lwt|) which also uses
\texttt{httpaf} and \texttt{libev}. Unlike using effect handlers, the Lwt
version is written in monadic style and does not have the notion of a thread
per request. For comparison, we include a Go 1.13 version (|go|) that uses the
\texttt{net/http}~\cite{nethttp} package. As both the OCaml versions are single
threaded, the Go benchmark is run with \texttt{GOMAXPROCS=1}.

\begin{figure}
	\begin{minipage}{0.49\linewidth}
		\centering
		\includegraphics[width=\linewidth]{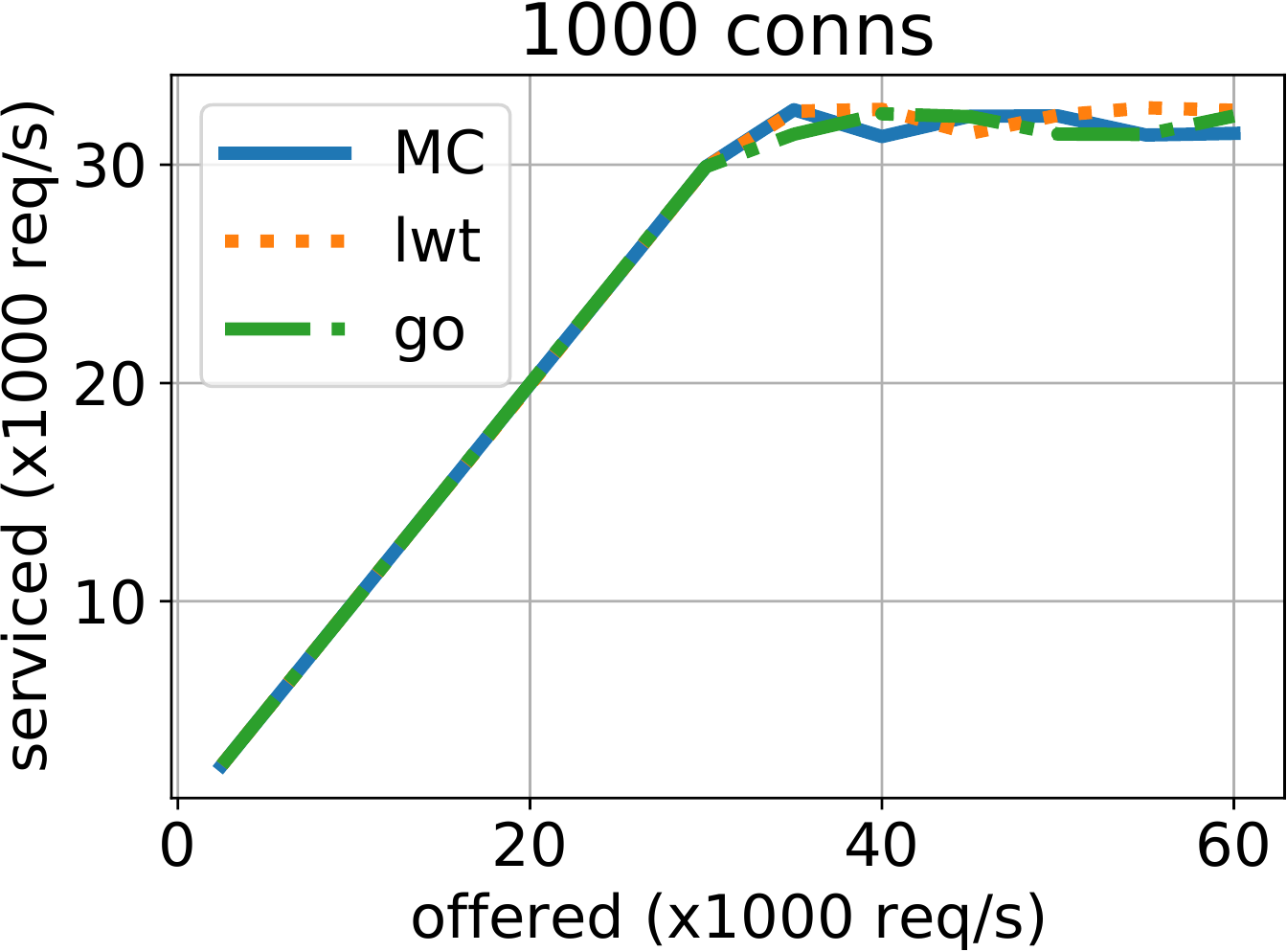}
		\subcaption{Throughput}
		\label{grf:throughput}
	\end{minipage}
	\begin{minipage}{0.49\linewidth}
		\includegraphics[width=\linewidth]{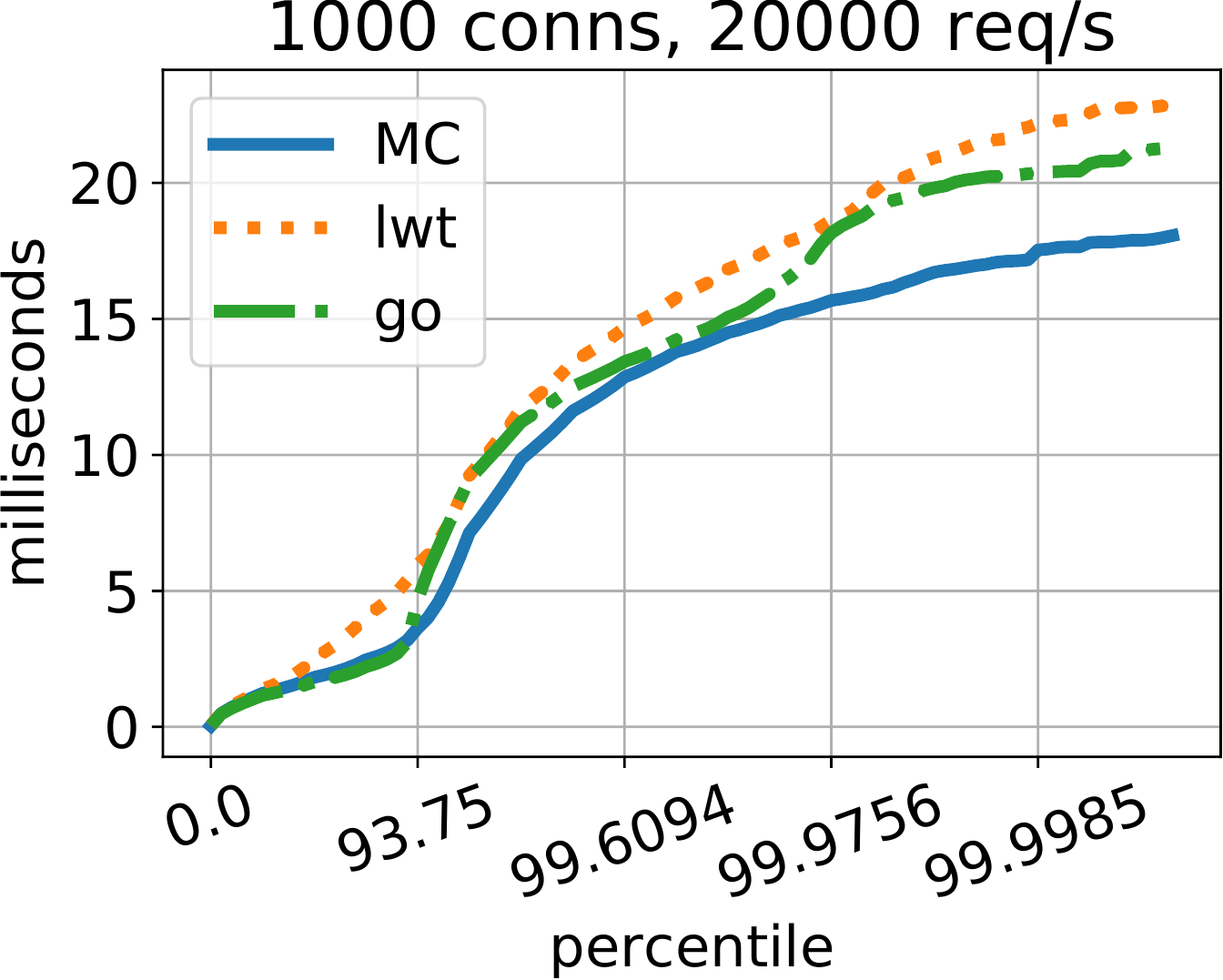}
		\subcaption{Tail latency}
		\label{grf:latency}
	\end{minipage}
	\vspace{-3mm}
	\caption{Web server performance.}
	\vspace{-5mm}
\end{figure}

The client workload was generated with \texttt{wrk2}~\cite{wrk2}. We maintain
1k open connections and perform requests for a static web page at different
rates, and record the service rate and latency. The throughput and tail latency
graphs are given in Figures~\ref{grf:throughput} and~\ref{grf:latency}. In all
the versions, the throughput plateaus at around 30k requests per second. We
measure the tail latencies at 2/3rd of this rate (20k requests per second) to
simulate optimal load. We observe that both of the OCaml versions remain
competitive with |go|, and |MC| performs best in terms of tail latency.

Multicore OCaml supports backtraces for continuations in addition to backtraces
of the current stack as in stock OCaml. Using effect handlers in a system such
as a web server aids debugging and profiling because it is possible to get a
backtrace snapshot of all current requests. This feature is available in
Go~\cite{gopprof}, but not in OCaml concurrency libraries such as Async and
Lwt which lack the notion of a thread.

\vspace{-2mm}
\section{Related Work}
\label{sec:related}

There are several strategies for implementing effect handlers. Eff~\cite{Eff},
Helium~\cite{Biernacki20}, Frank~\cite{Frank} and the Links server
backend~\cite{Hillerstrom20} use an interpreter similar to our operational
semantics to implement effect handlers. Effekt~\cite{Effekt}, Links JavaScript
backend~\cite{Hillerstrom20} and Koka~\cite{Leijen17} use type-directed
selective CPS translation. These language are equipped with an effect system,
which allows compiling pure code in direct style and effectful code in CPS.
Leijen~\cite{Leijen14} implements effect handlers as a library in C using stack
copying. C allows pointers into the stack, so care is taken to ensure that
when continuations are resumed, the constituent frames are restored to the same
memory addresses as at the time of capture. Kiselyov et al.~\cite{Kiselyov18}
use an implementation of multi-prompt delimited continuations as an OCaml
library~\cite{Kiselyov12} to embed the Eff language in OCaml. Indeed, Forster et
al.~\cite{Forster19} showed that in an untyped setting, effect handlers,
monadic reflection and delimited control can macro-express each other.

Multicore OCaml uses the call stack for implementing continuations (as
do~\cite{Leijen14, Kiselyov12}), but with one-shot continuations. Bruggeman et
al.~\cite{Bruggeman96} show how to implement one-shot continuations efficiently
using segmented stacks in Scheme. Farvardin et al.~\cite{Farvardin20} perform a
comprehensive evaluation of various implementation strategies for continuations
on modern hardware. Multicore OCaml stacks do not neatly fit the description of
one of these implementation strategies -- they are best described as using the
resize strategy from Farvardin et al. for each of the fibers, which are linked
to represent the current stack and the captured continuations. Kawahara et
al.~\cite{Kawahara20} implement one-shot effect handlers using coroutines as a
macro-expressible translation, and present an embedding in Lua and Ruby. Lua
provides asymmetric coroutines~\cite{Lua} where each coroutine uses its own
stack similar to how each handled computation runs in its own fiber in
Multicore OCaml.

Multicore OCaml is not the first language to support stack inspection in the
presence of non-local control operators. Chez Scheme supports continuation
marks~\cite{Flatt20} which permit stack inspection as a language feature. This
enables implementation of dynamic binding, exceptions, profilers, debuggers,
etc, in the presence of first-class continuations. As the authors note,
continuation marks can be implemented using effect handlers, but direct support
for continuation marks leads to better performance. In this work, we focus on
retaining the support for stack inspection through DWARF unwind tables in the
presence of effect handlers.

The interaction of non-local control flow and resources has been studied
extensively. Scheme uses |dynamic-wind|~\cite{R5RS}, which is a generalisation
of Common Lisp |unwind-protect|~\cite{Steele90}, which ensures de-allocation
and re-allocation of resources every time the non-local control leaves and
enters back into a context. |dynamic-wind| is not quite the right abstraction
as resources need to be cleaned up only on non-returning
exits~\cite{Kiselyov,Sitaram03}. This requires distinguishing returning exits
from non-returning ones.

Multicore OCaml builds on the existing defensive coding practices against
exceptions to clean up resources on non-returning exits. We assume that the
continuations are resumed exactly once using |continue| or |discontinue|. Under
this assumption, when a computation performs an effect, we expect the control
to return. For the non-returning cases (value and exceptional return), the code
already handles resource cleanup.

OCaml does not have a \texttt{\footnotesize try/finally} construct commonly
used for resource cleanup in many programming languages. The OCaml standard
library~\cite{FunProtect} as well as alternative standard libraries such as
Base~\cite{BaseProtect} and Core~\cite{CoreProtect} provide mechanisms
analogous to |unwind-protect|, which are in turn implemented using exception
handlers. Thus, the linear use of continuations enabled by the |discontinue|
primitive ensures backwards compatibility of legacy OCaml systems code under
non-local control flow introduced by effect handlers.

Leijen~\cite{Leijen18} explicitly extends effect handlers with |initially| and
|finally| clauses in Koka for resource safety. Dolan et al.~\cite{TFP17}
describe the interaction of effect handlers and asynchronous exceptions. This
is orthogonal to the contributions of this paper. Our focus is the compiler and
runtime system support for implementing effect handlers.

\vspace{-2mm}
\section{Conclusions}
\label{sec:conc}

Our design for effect handlers in OCaml walks the tightrope of maintaining
compatibility (for profiling, debuggers and minimal overheads for existing
programs), while unlocking the full power of non-local control flow constructs.
Our evaluation shows that we have achieved our goal: we retain compatibility
with a surprisingly low performance overhead for sequential code that preserves
the spirit of ``fast exceptions'' that has always characterised OCaml
programming. We believe that the introduction of effect handlers into OCaml
implemented using lightweight fibers, along with a parallel
runtime~\cite{Sivaramakrishnan20}, as has been demonstrated in our work, will
open OCaml to highly scalable concurrent and task-parallel applications with
minimal hit to sequential performance.

\section*{Acknowledgements}

We thank Sam Lindley, Franc\'{o}is Pottier, the PLDI reviewers and our
shepherd, Matthew Flatt, whose comments substantially helped improve the
presentation. This research was funded via Royal Commission for the Exhibition
of 1851 and Darwin College Research Fellowships, and by grants from Jane Street
and the Tezos Foundation.

\bibliographystyle{ACM-Reference-Format}
\balance
\bibliography{retro-concurrency}

\end{document}